\documentclass[12pt,preprint]{aastex}

\newcommand{\Lya}{\hbox{Ly$\alpha$}}

\newcommand{\Jup}{\hbox{$J^\prime$}}
\newcommand{\Jlo}{\hbox{$J^{\prime\prime}$}}
\newcommand{\Vup}{\hbox{$v^\prime$}}
\newcommand{\Vlo}{\hbox{$v^{\prime\prime}$}}
\newcommand{\Elo}{\hbox{$E^{\prime\prime}$}}
\newcommand{\kms}{\hbox{km s$^{-1}$}}
\newcommand{\IUE}{\textit{IUE}}

\newcommand{\HST}{\textit{HST}}
\newcommand{\STIS}{STIS}            % supposed to be roman typeface -JAV
\newcommand{\FUSE}{\textit{FUSE}}
\newcommand{\IDL}{IDL}              % assume roman typeface for now -JAV

\slugcomment{Accepted by ApJ}
\shorttitle{FUV Spectra of TW Hya.  II.  Models of H$_2$ Fluorescence}
\shortauthors{Herczeg et al.}

\begin{document}

% Title.
\title{The Far-Ultraviolet Spectra of TW Hya.  II.  Models of H$_2$ Fluorescence in a Disk$^1$}

% Authors.
\author{Gregory J. Herczeg, Brian E. Wood, and Jeffrey L. Linsky}
\affil{JILA, University of Colorado and NIST, Boulder, CO 80309--0440}
\email{gregoryh@origins.colorado.edu,woodb@marmoset.colorado.edu,jlinsky@jila.colorado.edu}

\author{Jeff A. Valenti}
\affil{Space Telescope Science Institute, Baltimore, MD 21218}
\email{valenti@stsci.edu}

\author{and Christopher M. Johns-Krull}
\affil{Department of Physics and Astronomy, Rice University, 6100 Main
       St. MS-108, Houston, TX 77005}
\email{cmj@rice.edu}

\footnotetext[1]{Based on observations with the NASA/ESA Hubble Space
Telescope, obtained at the Space Telescope Science Institute, which is
operated by the Association of Universities for Research in Astronomy,
Inc. under NASA contract NAS5-26555.  This work is also based on data
obtained for the FUSE Science Team by the NASA-CNES-CSA
\FUSE\ mission, operated by Johns Hopkins University.}

% Abstract
\begin{abstract}
We measure the temperature of warm gas at
planet-forming radii in the disk around the classical T Tauri star (CTTS) TW Hya by
modelling the H$_2$ fluorescence observed in \textit{HST}/STIS and
\textit{FUSE} 
spectra.  Strong \Lya\ emission irradiates a warm disk surface within 2 AU
of the central star and pumps certain excited levels of H$_2$.
We simulate a 1D plane-parallel atmosphere to
estimate fluxes for the 140 observed H$_2$ emission lines and 
to reconstruct the \Lya\ emission profile incident upon the warm
H$_2$.  The excitation of H$_2$ can be determined from relative line
strengths by measuring self-absorption in lines with low-energy lower
levels, or by reconstructing the
\Lya\ profile incident upon the warm H$_2$ using the total flux from a
single upper level and the opacity in the pumping transition.
Based on those diagnostics,
we estimate that the warm disk surface has a column density of $\log
N($H$_2)=18.5^{+1.2}_{-0.8}$, a temperature $T=2500^{+700}_{-500}$ 
K, and a filling factor of H$_2$, as seen by the source of \Lya\ emission, of
$0.25\pm0.08$ (all 2$\sigma$ error bars).
TW Hya produces approximately $10^{-3}$ $L_\odot$ in the
FUV, about 85\% of which is in the \Lya\ emission line.  From the
\ion{H}{1} absorption observed in 
the \Lya\ emission, we infer that dust extinction in our line 
of sight to TW Hya is negligible.
\end{abstract}

% Official ApJ keywords in alphabetical order.
% See http://www.journals.uchicago.edu/ApJ/information.html

% Citation commands:
%
% \citet{Nam98} for text citations, e.g. "Name (1998) says..."
% \citep{Nam98} for parenthetical citations, e.g. "is true (Name 1998)."
% \citep{Nam98,Alt01} to cite multiple refs, e.g. "(Name 1998; Alt 2001)"
% \citep{Nam98,Nam01} automatically merges, e.g. "(Name 1998; 2001)"
% \citet*{Nam98} or \citep* for the first citation of 3 author papers.
% \citet*{Nam98} or \citep* for the first citation of 3 author papers.
% \citep[e.g.,][]{Nam98} for leading text, e.g. "(e.g., Name 1998)"
% \citep[, for example]{Nam98} for trailing text.
% \citep[e.g.,][, for example]{Nam98} for leading and trailing text.
% \citealt{Nam98} is \citet{Nam98} without parentheses.
% \citealp{Nam98} is \citep{Nam98} without parentheses.
% \citeauthor{Nam98} only lists authors, e.g. "Name"
% \citeyear{Nam98} only lists the year, e.g. "1998"
% \citeyearpar{Nam98} only lists the year in parentheses, e.g. "(1998)"
% \citetext{\citealp{Nam98}; see also \citealp{Alt01}} embeds text, e.g.
%         "(Name 1998; see also Alt 2001)"
%
% Look up "natbib" on the web for a more detailed description.

%%%%%%%%%%%%%%%%%%%%%%%%%%%%%%%%%%%%%%%%%%%%%%%%%%%%%%%%%%%%%%

\section{INTRODUCTION}
Classical T Tauri Stars (CTTSs) are roughly solar-mass pre-main sequence
(PMS) stars that are accreting gas from their circumstellar disks.
The extent and mass of dust in disks have been
determined from IR spectral energy distributions (SEDs) 
and imaging 
\citep[see review by][]{Zuc01}.   Cold
gas at large radii is traced by observations of molecules such as CO
and HCN \citep[e.g.,][]{Aik02}.  Warmer neutral and ionized gases are used as
diagnostics of accretion \citep[e.g.,][]{Gom99,Joh02}.
However, observing the gas in the disk close to the star, which is
important for the formation and evolution of planets, has been
difficult.  Gas at these radii can induce planet migration \citep{Gol79},
dampen the eccentricity of terrestrial planets \citep{Agn02}, and is
neccessary for accretion onto giant planets \citep{Lis93}.  

In search of gas at planet-forming radii, \citet{Cal91} investigated the formation of CO
overtone lines at 2.3 $\mu$m in disks around CTTSs.  These models predict
that CTTSs
as a group should show both CO emission and absorption, while
Herbig Ae/Be stars should show CO in emission.
  Using high spectral resolution
observations, \citet{Car93} showed that the CO
overtone emission from the probable Herbig Ae/Be star WL 16 arises in the
inner disk, confirming the prediction of \citet{Cal91}.  \citet{Naj96}
 find the same result for the Herbig Ae/Be star
1548C27.  However, broader searches at high spectral resolution for CO
overtone emission or absorption from CTTSs generally find no evidence
for contributions to these lines from the disks around these 
stars \citep{Cas96,Joh01b}.
On the other hand, \citet*{Naj03} find
that CO fundamental emission near 4.6 - 4.9 $\mu$m is common among
CTTSs.  They suggest that overtone emission
  has not been detected at the observed CO temperatures (1100 --
  1300 K), mainly because the overtone bands have much lower
  transition rates.  However, CO is not expected to
be the dominant gas constituent in the disks around CTTSs, and it is the
major component of the gas, H$_2$,  that we seek to probe here.

Previous observations of 
H$_2$ have not provided a clear
picture of the gas in the disk.
\citet{Thi01} detected cold H$_2$ emission in the pure 
rotational S(0) and S(1) lines from ISO observations of a large sample of
CTTSs, Herbig Ae/Be stars, and debris-disk stars.
However, in ground-based observations of the S(1) and S(2) lines, using much
smaller apertures than ISO, \citet{Ric01} and \citet{She03} 
did not detect any H$_2$ emission from some of
the same sources, indicating that the emission is probably extended.
  On the other hand, the emission in the H$_2$ 1-0 S(1) transition detected by \citet*{Bar03} 
in Phoenix spectra of 4 T Tauri stars is narrow and centered at the radial
 velocity of the stars, presumably formed in low-velocity circumstellar gas.

While H$_2$ lines in the IR are typically weak and are observed 
against a strong dust
continuum, fluorescent H$_2$ lines dominate the far-ultraviolet (FUV)
spectrum of CTTSs at wavelengths longward of \Lya.  
\citet{Ard02} found that narrow H$_2$ lines can be blueshifted by up to 20
\kms\ in \HST/GHRS spectra of 
CTTSs, observed with a $2\arcsec\times 2\arcsec$ aperture.
High-resolution echelle and long-slit 
spectroscopy with \HST/STIS is providing the critical data needed to 
identify the source of the H$_2$ emission.
Observations of T Tau reveal two components of H$_2$ emission.  \citet{Sau03} 
and \citet{Wal03} detect off-source UV H$_2$ emission, that is pumped near
line-center of \Lya, and most likely produced where stellar outflows shock
the surrounding ambient molecular material.  The same set of fluorescent
H$_2$ lines are detected in low-excitation HH
objects such as HH43 and HH47 \citep{Sch83,Cur95}, suggestive of a similar
excitation mechanism.  
\citet{Wal03} find that the on-source H$_2$ emission 
is photoexcited by a much broader \Lya\ emission line, likely produced by 
the accretion shock at the surface of T Tau.

In \citet[][hereafter Paper I]{Her02}, we presented observations of H$_2$
fluorescence in the UV spectrum 
of TW Hya obtained with \HST/STIS and \FUSE.  Several lines of evidence
suggested a disk 
origin for the H$_2$ emission:
(i) the H$_2$ emission is not spatially extended beyond a point source in the
cross-dispersion direction, (ii) the H$_2$ lines are emitted interior to TW
Hya's wind because emission in one H$_2$ line is supressed by 
\ion{C}{2} wind absorption, (iii) the H$_2$ line centroids have the same
radial velocity as the photospheric lines of TW Hya, (iv) no H$_2$
absorption is detected against the 
\Lya\ emission, 
and (v) the TW Hya Association is isolated from large reservoirs 
of interstellar molecular material, making a circumstellar origin for the
H$_2$ emission unlikely.  Because the H$_2$ emission is not extended beyond 
a point source, the spatial resolution of \HST/STIS restricts the emitting
region to be within 2 AU of the star.

Analysis of on-source IR H$_2$ emission lines to date has been limited
because typically only one or two lines are observed.  The IR lines could
be excited by shocks, UV fluorescence, 
X-rays, or thermal
heating.  In contrast, over 140 H$_2$ lines from 19
different upper levels are detected in the FUV spectrum of TW Hya.  
The excitation of H$_2$ can be determined
  from relative line strengths in this rich spectrum either by
  measuring self-absorption in low excitation members of
  individual fluorescent progressions or by using an assumed
  Ly$\alpha$ profile to reconstruct initial level populations
  in various pumping transitions.

In this paper, we calculate the
temperature of warm gas at planet-forming distances
by modelling warm 
H$_2$ in a disk around TW Hya. 
We use the H$_2$ emission to reconstruct
  the Ly$\alpha$ profile incident on the fluoresced H$_2$. The
  observed Ly$\alpha$ profile differs from the incident profile
  because of interstellar absorption along our line of sight
  to TW Hya.  We discuss the strength of Ly$\alpha$ emission and its affect 
on the circumstellar disk.

%%%%%%%%%%%%%%%%%%%%%%%%%%%%%%%%%%%%%%%%%%%%%%%%%%%%%%%%%%%%%%%

\section{PROPERTIES OF TW HYA}
TW Hya is the namesake of the 10 Myr old TW Hya Association (TWA),
which most likely originated in the Sco-Cen OB association \citep*{Mam00}.
TW Hya is one of the oldest known stars still in the CTTS phase, and its mass
accretion rate of about $2-3\times10^{-9} M_\odot$ yr$^{-1}$ is lower than
the mass accretion rate of most of the 1--2 Myr old CTTSs that are found in
Taurus \citep{Ken95}.
From PMS evolutionary tracks, \citet{Web99}
estimate that TW Hya has a mass of $0.7$ $M_\odot$ and a radius of $1.0$
$R_\odot$.   
Gaseous material is often associated with young stars, but is not
found in the nearby TWA, so the extinction is low.  
The TWA is located at a distance of
approximately 56 pc \citep{Web99}, compared with 140 pc for Taurus.
Consequently, TW Hya is by far the brightest known CTTS in the UV and in
X-rays.  
Most of the UV radiation from TW Hya 
is presumably produced at the accretion shock.
The disk is viewed face-on
\citep[e.g.,][]{Zuc95,Ale02}, which prevents signficant
Keplerian broadening in the H$_2$ line profiles.  The disk 
mass is roughly $(1.5-3)\times10^{-2}$ $M_\odot$ \citep{Wil00,Tri01} and
extends more than 225 AU from the star \citep[e.g.][]{Kri00}.
Based on estimates of the stellar mass,
  radius, and rotation period of TW Hya, \citet{Joh01a} 
calculate that the disk at 6 $R_*$ corotates with the
  stellar surface. In the \citet{Shu94} model, the disk is
  truncated at corotation.
\citet{Wei02}
estimate from modelling the IR SED of TW
Hya that the dust in the disk is
truncated at about 11 $R_*$.  In their models of the IR SED, 
\citet{Cal02} found an underabundance 
micron-sized dust grains within 4 AU of
TW Hya.  Figure 1 shows a schematic model of TW Hya and its disk.  

Since the mass accretion rate of CTTSs can vary, we calculate
the mass accretion rate at the epoch when we observed TW Hya.  We then
proceed to estimate the extinction using \ion{H}{1} absorption against
\Lya\ emission and the SED of the NUV continuum that is produced in
an accretion shock.  
The observations discussed in this paper were described in Paper I and are
listed in Table 1.

\subsection{Mass Accretion Rate}
The strong excess blue and UV emission from CTTSs can
be modelled as emission from shocks at accretion footpoints on the star,
 which leads to
estimates of the mass accretion rate.
We estimate the accretion rate of TW Hya by following the method of
\citet*{Val93}, who modelled observed blue spectra with the sum of a
  photospheric template and emission from a slab of hot
  isothermal hydrogen to crudely simulate emission from the
  accretion shock.

We use the HST/STIS spectrum of the K7 weak-lined TTS V819 Tau (see Table 1), 
spanning 1100--6000 \AA,
 as a non-accreting template for the photospheric emission from TW Hya.  
For V819 Tau we adopt a visual extinction
  of $A_V=1.52$, which is the average of values determined by
\citet{Ken95} and \citet{Whi01}.   We then
  use the interstellar extinction law of \citet{Car89}
to remove the effects of extinction from the
  observed spectrum of V819 Tau.
The NUV and FUV
emission from V819 Tau is much fainter than that of TW Hya, 
and is not significant in this analysis.

The spectrum of the model slab is determined by the slab 
temperature $T$, density $n$,
thickness $l$, extinction $A_V$ towards the slab, and a slab surface area parameter 
$\delta$, as described in \citet{Val93}.
The slab surface area parameter $\delta$ corresponds to the fraction
of the stellar surface area covered by accretion-related emission, 
and is calculated from the scale factor applied to the slab spectrum.  The
scaling 
factor between the non-accreting template star and TW Hya is another 
free parameter.  We find the best-fit model parameters by fitting the
synthetic slab spectrum plus the scaled template spectrum to the NUV and optical spectrum of TW Hya, minimizing
$\chi^2$ using the {\it
amoeba} function in {\IDL}, which is based on the downhill simplex
minimization method.
The best-fit parameters for the 
slab are $A_V=0.0$, $T=9900$ K, $n=1.7\times10^{14}$ cm$^{-3}$, $l=125$ km,
$\delta=0.008$, and a template scaling factor of 1.18.

Figure 2 shows our fit to the optical spectrum of TW Hya, after subtracting
the scaled template spectrum of V819 Tau.  The accretion
continuum calculated in this model is consistent with the observed NUV
continuum.  However, the observed continuum flux rises
shortward of 1700 \AA,
which could result from (i) an H$_2$ dissociation continuum produced by
electron collisions \citep{Liu96}, (ii) H$_2$ fluorescence pumped by the FUV
continuum, or (iii) by a hot accretion or activity
component.  The SED of the FUV continuum is similar to the continuum
observed in HH2, which is probably produced by H$_2$ dissociation
\citep{Ray97}.  However, the FUV continuum of TW Hya appears smooth and does
not show strong H$_2$ emission lines at 1054 \AA\ and 1101 \AA, which are
detected towards HH2.

Assuming that half of the potential energy of the accreting gas is converted
into hydrogen continuum emission$^2$, then the mass accretion rate of TW Hya can be estimated by
\begin{equation}
\dot{M}_{*}=\frac{1.25R_*L_{\rm slab}}{GM_*},
\end{equation}
where $L_{\rm slab}=3.5\times10^{-10}$ erg cm$^{-2}$ s$^{-1}$ is the 
slab flux integrated over all
wavelengths.  The leading factor of 1.25, derived by \citet{Gul98} for a
magnetospheric accretion geometry, replaces the factor of 2 from a spherical
accretion geometry.
We adopt $0.7~M_\odot$ and $1.0~R_\odot$
for the mass and radius of TW Hya \citep{Web99}.  
Uncertainties in the stellar mass,
radius, and distance are all of order 20\%.  We calculate an accretion rate
onto the star of $\dot{M}\approx 2\times10^{-9}$ $M_\odot$ yr$^{-1}$, which is consistent 
with the accretion
rate of $2\times10^{-9}$ $M_\odot$ yr$^{-1}$ estimated by \citet{Ale02}
using a similar 
method, and is larger than the accretion rate of $5\times10^{-10}$ $M_\odot$
yr$^{-1}$ estimated by \citet{Muz00} using an analysis of the H$\alpha$
line.
\footnotetext[2]{See \citet{Lyn74} and \citet{Har91} for details.}

\subsection{Extinction to TW Hya}
In this section, we argue that the
extinction to TW Hya is negligible by measuring the hydrogen column
density directly from the shape of the observed \Lya\ line, 
and then using an ISM extinction  law to infer the dust extinction.
This argument does not
necessarily imply that the extinction towards the warm H$_2$ is negligible,
because some dust could be mixed with the warm H$_2$, leading to extinction of
H$_2$ fluxes without any extinction of the stellar or accretion emission.
We will discuss this
possibility in \S 5.

Figure 3 shows the observed \Lya\ profile, which is likely
produced in an accretion column, contains a dark, broad absorption feature that
extends from +200 \kms\ to -500 \kms.
The absorption is produced by \ion{H}{1} in the wind and in
interstellar and circumstellar material.
Because circumstellar and interstellar
absorption are indistinguishable in our data, 
we model these two features with a single absorption feature.
All UV spectra of CTTSs show similar wind absorption in Mg II \citep{Ard02b}, so the star must be occulted by the outflow, regardless of inclination angle.

In the following analysis, we fit two Gaussian emission profiles and
either one or
two Voigt
absorption profiles to the observed \Lya\ profile.  
The intrinsic emission
profile is assumed to be a double-Gaussian profile, and the parameters of
the Gaussians are allowed to 
vary to best fit the observed profile when combined with the
absorption.
Based on the widely separated wind and interstellar 
absorption features observed in the 
\ion{O}{1}~1302~\AA,
\ion{C}{2}~1334~\AA, and \ion{Mg}{2}~2795~\AA\ lines, the interstellar and
wind absorption components towards TW Hya must be centered at $v=0$ and
$-130$ \kms, respectively.
For the interstellar component we assume a Doppler parameter of 
$b=11$ \kms, which corresponds to a
temperature of  7000~K that is typical of the local (warm) ISM \citep{Red00}.  An
interstellar column density is then determined by fitting the
absorption longward of line center, where any wind absorption should be
negligible.
 The wind absorption
extends to a
velocity of at least -500 \kms, and likely has a large Doppler parameter due to
velocity gradients along our line of sight. Given these assumptions, the
interstellar neutral hydrogen column density in our line of sight is  $\log
N($\ion{H}{1}$)=19.05\pm0.15$ (Fig. 3).
The upper limit of the column density in the wind is
$N($\ion{H}{1}$)<19.4$ for Doppler
parameters $b>25$ \kms

In a more conservative test to find the upper limit to the hydrogen column
density, we fit the \Lya\ absorption with a single component,
effectively combining any interstellar and wind absorption.
We vary the central wavelength of the absorption  
and set a low Doppler parameter
parameter, so that the damping wings generate the 
width of the absorption profile.  Figure 4 shows that the maximum possible
column density consistent with the data is $\log N($\ion{H}{1}$)=19.75$.  

The conservative upper limit of the hydrogen column density towards 
TW Hya is therefore $\log N($H$)<19.75$.  In the only
previous measurement of $N$(\ion{H}{1}) to TW Hya, \citet{Kas99} used
X-ray spectral energy distributions from {\it ROSAT} and {\it ASCA} to
estimate higher column densities of
$\log N($\ion{H}{1}$)=20.5\pm0.4$ and $21.4\pm0.1$, respectively.  
However, the X-ray analyses are highly reliant on the accuracy of plasma codes.
\citet{Kas99} estimate the hydrogen column density
by assuming solar abundances and using a two-temperature
emission
measure rather than a differential emission measure distribution. 
However, in their analysis of the {\it Chandra} X-ray spectrum of TW Hya,
\citet{Kas02} found that the abundances show an
inverse-FIP effect.  We further note that the sensitivity and flux
calibration of {\it ROSAT} and {\it ASCA} are relatively poor at soft
energies, where absorption from \ion{H}{1} ionization is measured.  Figure 4
shows  that the absorption calculated for the \Lya\ line for
$\log N($\ion{H}{1}$)=20.5$ is signficantly broader than the observed
absorption profile.  Our comparison of the calculated transmission function
with the observed \Lya\ profile provides a much simpler and more direct 
test of hydrogen 
column density than the X-ray analyses.
The discrepancy between the UV and X-ray measurements of N(H) 
could in principle be due to a variable circumstellar absorber, 
which \citet{Kas99} invoked to explain the order of magnitude 
discrepancy in calculations of N(H) from {\it ASCA} and {\it ROSAT}.
However, the many ultraviolet observations of TW Hya by \IUE, 
\HST/STIS, and \FUSE, which are widely spaced in time, show 
no evidence for large fluctuations in extinction.

We estimate an H$_2$ column density in our line of sight to TW Hya with \FUSE, as the bandpass contains many H$_2$ transitions from low rotational
levels of the ground electronic state, which are often seen in absorption
due to the ISM at temperatures typically $\sim 100$ K 
\citep{Rac02}.  H$_2$ absorption lines from
$\Vlo=0, \Jlo=0,1,2,3,5$ can in principle be 
observed against O VI emission \citep[e.g.][]{Rob01}.  
The \ion{O}{6} profiles are noisy (see Fig. 5), but we estimate an upper
limit of $\log N($H$_2)<18$ by calculating the transmission percentage for a 
range of temperatures and 
column densities, meaning the molecular fraction of H towards TW Hya is very low.

When we convert total hydrogen column density, 
$N$(\ion{H}{1})+$2N($H$_2)=N($H),
 to a reddening using the interstellar 
relationship of \citet*{Boh78},
\begin{equation}
\frac{N({\rm H I}+{\rm H}_2)}{E(B-V)}=5.8 \times 10^{21} {\rm cm}^{-2},
\end{equation}
we find that the reddening toward TW Hya is
$E(B-V)=A_V/R_V< 0.01$, where $R_V$ is a dimensionless
parameter that ranges between 2.5--6, depending on the size distribution of dust grains.
Significant deviations from the assumed dust-to-gas ratio
can occur in star-forming regions such as $\rho$ Oph, in which the dust
grains are likely larger than they are in the ISM.  However, the low
molecular fraction towards TW Hya is consistent with a low extinction
\citep{Sav77,Rac02}, and is not consistent with material found towards star-forming regions.
We conclude that this relationship
provides a reasonable estimate of the extinction along our sightline to
TW Hya.
Even if the dust-to-gas ratio along the line of sight to TW Hya were
  for some reason enhanced by an order of magnitude, then dust 
extinction would reduce the brightness of TW Hya by a factor of 2 
and have only a 20\% effect on the relative fluxes between 1200 \AA\ and 1650
\AA.

Inspection of the NUV continuum (Fig. 6) confirms the small
 extinction to TW Hya.  If we
assume that the NUV continuum is produced in an isothermal 
slab, as modelled in \S 2.1, then it can be well fitted with a hydrogen 
continuum only for extinctions $A_V<0.2$.
As noted in \S2.1, an accretion continuum best fits the NUV continuum for $A_V=0.0$.

%%%%%%%%%%%%%%%%%%%%%%%%%%%%%%%%%%%%%%%%%%%%%%%%%%%%%%%%%%%%%%%

\section{MODELLING H$_2$ LINE FLUXES}

%%%%%%%%%%%%%%%%%%%%%%%%%%%%%%%%%%%%%%%%%%%%%%%%%%%%%%%%%%%%%%%

In Paper I, we measured a total line emission 
flux of $1.94 \times 10^{-12}$ erg
s$^{-1}$ cm$^{-2}$ by summing 146 H$_2$ emission lines in the 19
progressions (the set of {\it R} and {\it P} transitions from a common upper
level) listed in Table 2.  Most of these lines are Lyman-band 
transitions fluoresced by the strong, broad \Lya\ emission
line.

\citet*{Woo02} discovered in their analysis of \HST/STIS\ observations of
Mira B that H$_2$ fluxes weaken towards short
wavelengths because of increasing H$_2$ line opacities.
This occurs because shorter wavelength lines are transitions to low energy levels, which are more populated than higher levels.
Figure 7 demonstrates that the same
effect occurs in the H$_2$ fluorescence sequences detected towards TW Hya.  In most sequences, fluxes
in many H$_2$ emission lines at short wavelengths are weaker than predicted from
theoretical branching ratios because of significant absorption in the same transitions.
Transitions from levels with low excitation energies, such as \Vup=0,
\Jup=1, have significant optical depth below 1300~\AA, but this does not happen 
for transitions from levels with large
excitation energies such as \Vup=4, \Jup=18 (see Fig. 7).  The \Vup=4,
\Jup=18 level does not have any downward transitions with significant
opacity in the \STIS\ bandpass. 

Following the Monte Carlo method of \citet{Woo02},
we compute H$_2$ opacities by modelling
H$_2$ fluorescence in a 1D isothermal plane-parallel 
atmosphere of H$_2$ irradiated by \Lya\ photons. 
We assume complete frequency and angular 
redistribution when photons scatter in the slab.
Forty-one wavelength grid points across each H$_2$ line profile 
track differences in escape
  probability across the line profile because
because a photon that 
scatters into the wings of a line escapes from a slab more easily 
than a photon 
at line center.  A \Lya\ photon enters the slab at an angle $\theta$ with
respect to the slab surface normal and photoexcites an H$_2$
molecule.  The H$_2$ molecule quickly de-excites to the ground electronic
state through one
of the many available routes, or dissociates with a
 probability $P_{Dis}$ [listed in Table 2, calculated by \citet*{Abg00}].
Depending on the optical depth
of H$_2$ in the lower level of the downward transition, the resulting
photon may 
be temporarily absorbed by another H$_2$ molecule or may emerge from either
side of the 
slab at an angle
$\phi$ with respect to the normal.  In our simple model, H$_2$ 
line photons created fluorescently in the slab either escape, 
are converted to another H$_2$ line in the same progression, or dissociate
an H$_2$ molecule.  No other photon processes are considered.  We tabulate
photons that exit the slab on
the \Lya\ entry side (reflected photons) and opposite side (transmitted
photons) using 40 bins of equal solid angle on each side of the slab.
We refer the reader to \citet{Woo02} for further details 
of this Monte Carlo code.

We use the term ``model'' to refer to the simulated plane-parallel
atmosphere described in the preceding paragraph, and the term ``geometry''
to refer to the morphology of the H$_2$ around the \Lya\ emission source.  
In the context of the model, the H$_2$ line fluxes depend on the
temperature and column density of the slab, the angle of
incidence $\theta$ of the Ly$\alpha$ photon into the disk, the outward
angle $\phi$ of
the emerging photon, the side of the slab from which the photon
escapes, and the optical depth of the transition.  We assume that extinction is
negligible in the slab and in any interstellar material.   Figure 8 shows two
possible geometries that we use to approximate 
a thick disk and a thin disk.  The term ``thick disk'' applies to a
star-disk geometry in which
most of the H$_2$ illuminated by \Lya\ is at the inner edge of the disk.  The term ``thin disk'' applies to a geometry
in which most of the H$_2$ illuminated by \Lya\ is at the disk surface.
In this section 
and \S 4, we present detailed
results for the thick disk geometry that reproduces the 
data well.  In this adopted geometry, Ly$\alpha$ photons enter the slab
normally
($\theta=0^\circ$), and the H$_2$ photons that we observe emerge at an angle
$\phi=76^\circ-79^\circ$ to the normal.
This choice is intended to present
one approximation of a plausible disk geometry for which we can present
detailed results, and
is not meant to exclude other geometries.
In \S 5, we present
general results for many geometries to demonstrate that our primary result, the
temperature of the H$_2$ gas, is relatively insensitive 
to geometries, including those that are significantly different from the adopted geometry.

We calculate the relative flux of each
transition within a single progression, given  
an H$_2$ excitation temperature 
and column density of 
the slab, and a separate flux normalization factor for each progression.
Models are calculated for temperatures in the range
$T=$1000--5500~K, sampled at 100~K intervals, and column
densities (in cgs
units) of
$\log N($H$_2$)=15--23, sampled at 0.1~dex intervals.  
$N$(H$_2$) and $T$ are then determined by fitting
simultaneously the relative line fluxes for 19 observed
progressions.
Five lines
are not used because they are
blended, and one line is not used because 
it is anomalously weak due to \ion{C}{2}
wind absorption (see Paper I).  The \FUSE\ lines, observed six weeks after the \STIS\
observations,  are not used in 
building these models due to the potential problems with flux
variability and cross-calibration between \STIS\ and \FUSE.  The fits have
21 variables:  19 normalization factors, $N($H$_2)$, 
and $T$.  In typical models, 46 of the 140 H$_2$ line fluxes are 
sensitive to H$_2$ line opacity, and hence $N($H$_2$) and $T$.  The 
other 94 lines, which have only a weak dependence on $N($H$_2$) 
and $T$, determine the 19 normalization factors.  We calculate 
$\chi^2_{\rm H2}$
 for only the 46 lines which depend on $N($H$_2$) and $T$.  
Confidence contours of 
$1-5\sigma$, calculated from values of
$\Delta\chi^2_{\nu, {\rm H2}}$ for 44 degrees of freedom, are shown in Figure 9, with a minimum 
$\chi^2_{\nu, {\rm H2}}=2.22$ at $\log N($H$_2)=17.9$ and $T=2600$ K.

Based on this analysis, and additional constraints provided by the \Lya\
profile analysis described in \S4, we ultimately decide a thick disk model
with $T=2500$ K and $\log  N$(H$_2)=18.5$ is our best model, which we adopt
for the analysis (see \S 6.1).  Figure 10 compares the observed
spectrum compared with a synthetic spectrum computed from this model.
The adopted model correctly predicts the fluxes of all strong H$_2$ lines.  
Only one observed line flux differs significantly from the model flux.  The
model predicts a flux of
$46 \times 10^{-15}$ erg cm$^{-2}$ s$^{-1}$ in the 0-4 R(1)
transition at 1333.797~\AA.  The measured flux in this line,
$(7.9\pm0.7)\times 10^{-15}$ erg cm$^{-2}$ s$^{-1}$, is significantly
reduced by \ion{C}{2} wind absorption, as
discussed in Paper I.  
The 0-3 P(3) transition at 1283.16 \AA, with a measured flux of
$28.0\times10^{-15}$~erg~cm$^{-2}$~s$^{-1}$, is moderately weaker than the
model flux of $34.8 \times0^{-15}$
erg cm$^{-2}$ s$^{-1}$, possibly because of H$_2$ absorption in the 2-2
R(15) transition, centered 8 \kms\ from the 0-3 P(3) transition.
Table 3 lists predicted fluxes 
for Lyman-band H$_2$ transitions in the \FUSE\ bandpass.  Future
 measurements of
flux in these transitions would provide a good test for our model.  
Lines at $\lambda<1200$ provide particularly good constraints since they are transitions to very low energy levels, and as a result have large opacities.

Neither this model nor the alternative models presented in \S 5 rigorously simulate a
disk geometry.  Because the structure of disks is uncertain near the
truncation radii,  
we do not have an
accurate picture of the 
geometry of the H$_2$ emission region.  Without 
this {\it a priori} knowledge, the thick disk geometry described here
is not clearly better or worse than any other oversimplified geometry that 
we might assume.  In \S 5, we will show that the temperature of the
molecular region is relatively insensitive to geometry, and we will rule
out certain geometries that cannot sufficiently explain 
the observed H$_2$ spectrum.  Since our models make as few assumptions as
possible, the primary result of this paper, the
temperature of the H2 gas, may apply to a wide
range of other morphologies.

%%%%%%%%%%%%%%%%%%%%%%%%%%%%%%%%%%%%%%%%%%%%%%%%%%%%%%%%%%%%%%%

\section{RECONSTRUCTED \Lya\ PROFILE}
In this section,
we reconstruct the \Lya\ profile seen by the warm H$_2$ gas at the 18
pumping wavelengths between 1212--1220 \AA.
Fluorescent H$_2$ emission occurs because certain Lyman-band
transitions, with lower levels $1-2$ eV above the ground state, 
have wavelength coincidences with the broad \Lya\ profile.
  The emission in each 
progression depends on
the \Lya\ flux at the pumping wavelength of the fluorescent transition,
the upward transition probability,
the opacity in the pumping transition, 
and  the filling factor of H$_2$ as seen from the source of the \Lya\ emission.
We reconstruct the incident \Lya\ profile following a procedure similar to
that used
by \citet{Woo02} to analyze fluorescent H$_2$ emission in the \HST/\STIS\
spectrum of Mira B.  A comparison of the reconstructed and observed \Lya\
profiles
provide another constraint on the values of $T$ and $N$(H$_2$),
that is independent of
the constraints provided by the analysis presented in \S 3.  The analysis
in \S 3 relied upon using the H$_2$ flux ratios within each fluorescence
progression to constrain $T$ and $N($H$_2)$, whereas the \Lya\ profile
reconstruction analysis presented in this section relies only on the total
flux from each upper level.

We calculate the optical depth in each H$_2$ pumping transition by assuming
that the absorption line profiles in the model 
slab are Voigt functions with a Doppler $b$ parameter, which depends
on an assumed temperature, and a damping parameter $a$ from \citet{Abg93}.
The total \Lya\ flux, $F_{abs}$, absorbed by a
given lower level is then
\begin{equation}
F_{abs}=\eta \times F_{inc}\times \int_\lambda 1-e^{-\tau_\lambda [T,N(H_2)]}{\rm d}\lambda
\end{equation}
where $F_{inc}$ is the incident \Lya\ flux per unit wavelength and  $\int_\lambda
1-\exp^{-\tau_\lambda [T,N(H_2)]}$d$\lambda$ is the equivalent width of the absorption profile
within the slab (see EQW$_{slab}$ in Table 2).  
Not every absorbed \Lya\ photon produces an observed H$_2$
photon. For the most highly excited states, the dissociation probability
($P_{dis}$ in Table 2) can be as large as 42\%. Some fraction
($R_{unseen}$ in Table 2) of the
emitted photons will be in spectral lines that we did not observe
due to line blends, noisy data, or insufficient wavelength coverage.
In our plane-parallel slab model, an emerging
photon travelling normal to the slab
sees a smaller optical depth than a photon travelling nearly parallel to
the slab.  As a result, the emitted flux in these models is largest for
angles normal to the slab.

For comparison with the observed \Lya\ profile, we
  compute reconstructed Ly-alpha profiles for the same 
  grid of $T$ and $N($H$_2$) values used in \S 3.
Model fluxes at 8 pumping transitions, all on the red wing of the
\Lya\ line, are fitted to 
the observed profile.  The observed \Lya\ flux at other pumping
wavelengths is corrupted by interstellar or wind absorption.
Two progressions, 3-3~R(2) at 1217.031~\AA\ and 3-3~P(1) at
1217.038~\AA, are not included in the fit because their absorption
profiles overlap, complicating the analysis.
Photoexcitation of H$_2$ by \ion{C}{3} emission via 
the transition 3-2~P(4) at 1174.923~\AA\ may also add flux to
the progression pumped by 3-3~R(2).
We judge the suitability of a model using a modified $\chi^2$ analysis to
measure the scatter in the red wing of the reconstructed profile.  
Errors in the \Lya\ line wing tend to be multiplicative, rather than
additive, so we define $\chi^2_{Ly\alpha}$
%%%
\begin{equation}
\chi^2_{Ly\alpha}=\sum (\frac{F_{obs}}{\sigma_{obs}}\log
\frac{F_{obs}}{F_{mod}})^2
\end{equation}
%%%
based on standard error propagation techniques from \citet{Bev69}.  Using
this formula, an overestimate and an underestimate of the reconstructed
\Lya\ profile by a
factor of 10 will result in the same contribution to $\chi^2_{Ly\alpha}$.

Figure 11 
shows confidence contours of $1-5\sigma$ calculated from 
$\Delta \chi^2_{\nu,Ly\alpha}$ for fits to the observed \Lya\ profile, 
with 5 degrees of freedom.
  In the thick disk geometry, a minimum $\chi^2_{\nu,Ly\alpha}=2.31$ is
obtained when comparing the reconstructed and observed \Lya\ profiles for 
$\log N$(H$_2$)=18.5, $T=2500$
K, and $\eta=0.25$.  The top panel of Fig. 12 shows that 
these parameters yield an excellent fit to the observed profile.
These parameters also reproduce the individual line fluxes 
well (see \S 3 and Fig. 9).  Since the confidence contours
calculated from the \Lya\ profile analysis place
a more stringent constraint on the temperature and column density than do
the contours from fitting individual line fluxes (see Fig. 11), we choose
on the values of $T$ and $N$(H$_2$) quoted above as our best estimates of these
parameters.  The model dependent quantities listed in Table 2 are all
calculated using these parameters.

More than 500 other possible upward transitions in the Lyman-band
have wavelengths
between 1210--1220 \AA.  Fluorescent emission in lines from these weaker
progressions is below our detection limit.
Table 4 identifies undetected progressions
with the largest predicted emission line fluxes. The
table includes our observational flux limit $F_{max}$
on the strongest line in the progression at wavelength
$\lambda_{max}$.  For any H$_2$ line that is not detected and
does not appear in Table 4, we place an upper flux limit
of $5 \times 10^{-15}$ erg cm$^{-2}$ s$^{-1}$.
We then use these upper limits to calculate an
upper limit of \Lya\ at the pumping wavelength to test the self-consistency
of our model.

Figure 12 shows three examples of reconstructed \Lya\ profiles (circles),
with upper limits to the \Lya\ flux at certain wavelengths calculated
from upper limits of lines in the undetected progressions (arrows), as described above.   The
top panel shows the \Lya\ profile constructed using the parameters for our
adopted model [$\log N($H$_2$)=18.5, $T=2500$~K, and $\eta=0.25$].  This
profile is smooth, and its wings correspond well with the wings of the
observed \Lya\ profile.
The upper limits on the flux of \Lya\ calculated from 
undetected progressions are all consistent with the \Lya\ profile
reconstructed from observed progressions.
The middle panel shows a \Lya\ profile calculated from a model with $\log
N($H$_2$)=19.0, $T=2000$~K, and $\eta=0.4$, which is
inconsistent with the data because of a large amount of scatter in the red wing
of the reconstructed
\Lya\ profile.  The bottom
 plot shows a \Lya\ profile calculated from a model with $\log N($H$_2$)=19.0,
$T=3000$~K, 
and $\eta=0.2$, which is also inconsistent with the data because 
at 1214.421~\AA\ the predicted \Lya\ emission
required to pump the $\Vup=7, \Jup=4$ level is significantly 
below the observationally constrained
flux at 1214.465 \AA.  Figure 13 is a reproduction of Figure 11, with 
dotted and diagonal lines added to show the
region of parameter space that are excluded because upper
limits on H$_2$ emission would imply a \Lya\ pumping flux
significantly lower than the nearby \Lya\ pumping flux for
a progression of observed H$_2$ lines. Parameter space
constraints are shown separately for upper limits on
progressions pumped via 7-4 P(5) at 1214.421 \AA\ alone
(dots) and via a set of others upward transitions (diagonal lines).
The parameter space
with $\eta>1$ (dashed lines in Fig. 13) are unphysical models, and are also
ruled out.

The three reconstructed \Lya\ profiles shown in Figure 12 are roughly similar 
to each other and
to most other \Lya\ profiles that we have reconstructed for other values of
$T$ and $N$(H$_2$).  
Each profile shows a clear
depression near \Lya\ line center that is not nearly as wide as
the observed \ion{H}{1} absorption.  
Figure 14 demonstrates that the 
absorption feature in the reconstructed \Lya\ profile can be characterized
by a column density 
$\log N($\ion{H}{1}$)\approx 18.7$ centered 90 \kms\ shortward of 
line center.
This \ion{H}{1} between the source of \Lya\
emissio and the warm H$_2$
could be in the base of a wind or at the disk surface.
Alternatively, the shape of the
\Lya\ profile could be an intrinsic self-reversal, in which case the
intervening \ion{H}{1} need not exist.
The blue side of the reconstructed 
\Lya\ profile is enhanced relative to the observed \Lya\ profile, 
because wind absorption in our line of sight to the star attenuates the
intrinsic emission.
This result supports our
conclusion from Paper I that, in our line of sight, the 
H$_2$ must be photoexcited interior to most of the 
wind that we detect.

%%%%%%%%%%%%%%%%%%%%%%%%%%%%%%%%%%%%%%%%%%%%%%%%%%%%%%%%%%%%%%%

\section{ALTERNATIVE GEOMETRIES}
In \S 3 and \S 4, we described in detail results for a dust-free 
thick disk geometry, which may not represent the true morphology of warm 
gas around the star.
In this section, we explore the same parameter space for a thin disk geometry,
shown schematically in Figure 8, and a cloud geometry, which
approximates H$_2$ emission from a cloud in our line of sight to the star.

The computed H$_2$ fluxes depend on the angle at which \Lya\ photons enter the 
disk, the exit angle of H$_2$ photons, and the extinction either 
within the disk or in the ISM.
Figures 15 and 16 and Table 5 present the best-fit parameters from fitting 
the H$_2$ line opacities for each of three geometries as a function of
the exit angle of H$_2$ and extinction within the disk.  
In general, while changes in the geometry of the system can mildly change
some results, 
the calculated temperature of the warm gas depends primarily upon 
the molecular physics and not on the geometry.  
The calculated column density of H$_2$ and its
physical interpretation, however, do depend on the assumed geometry.  In what
follows, the column density represents the amount of H$_2$ in a line of
sight normal to the disk surface, regardless of the path length the \Lya\
photons traverse through the disk.

\subsection{Thin Disk Geometry}
A thin disk (see Fig. 8) is characterized here as a disk for which the
filling factor of H$_2$ around the \Lya\ source is dominated by the surface
layers of the disk, rather than by the inner edge of the disk.  We approximate
this geometry with a model in which the incident \Lya\
photons enter a
plane-parallel slab at a large angle with respect to the disk normal and
are reprocessed into H$_2$
photons that emerge normal to the slab.
Because \Lya\ 
photons enter the slab 
at a large angle, the radiation is reprocessed into H$_2$ emission 
at a shallow depth, and the
 emerging H$_2$ photons see a smaller optical depth in this 
geometry than in the thick disk geometry.
For the thin disk geometry, with incident photons entering the disk at
$\theta=80^\circ$, the best-fit parameters are $T=2500$ K, $N($H$_2$)=20.0,
but with an unacceptable $\chi_{\nu,H2}^2=7.9$, for fits to H$_2$ line fluxes.
For fits of the reconstructed to the observed \Lya\ line, the confidence
contours move 
to smaller column densities by a factor of $\sin \theta$.
As the angle of incidence 
asymptotically approaches 90$^\circ$, the \Lya\ photons are converted into
H$_2$ photons close to the disk surface.
Since the optical depth traversed by the emitted H$_2$ photons is 
negligible,
the relative line fluxes with each progression differ from predicted
branching ratios only because of extinction, which we find to be unlikely
(see \S2.2 and \S5.3). 

\subsection{Cloud Geometry}
We next consider a geometry 
in which the H$_2$ is located between us and the source of \Lya\ photons.
In this geometry, the escaping photons must travel 
through a greater optical depth of H$_2$ than they would in the other geometries that we considered.
As a result, the acceptable
contours for fits to H$_2$ line fluxes move towards smaller column 
densities relative to the thick or thin disk geometries.
Even in this very different geometry, 
the confidence contours for the \Lya\ reconstruction fitting do
not move much in parameter space,
and they restrict the gas temperature to 2000--3000~K.
The best-fit parameters in fitting H$_2$ fluxes for this geometry are
$T=2600$ K, $N($H$_2$)=17.8, with $\chi^2_{\nu,H2}=2.8$.
The transmission geometry is probably not realistic for TW Hya, but it can
be used to set limits on the column of warm H$_2$ between us and the source of
\Lya\ emission.
The absence of any observed H$_2$ absorption in the \Lya\ profile (Fig. 3)
allows us to place a limit on the column density in this level 
of $\log N(\Vlo=2,\Jlo=1)<14.4$ and a total
column density of warm H$_2$ of $\log N($H$_2)<16.1$.  Thus, the detected
H$_2$ fluorescence does not occur in our line of sight to TW Hya.

\subsection{Dust Extinction in the Disk and the ISM}
In
\S2, we presented evidence that the dust extinction along our line of
sight to the \Lya\ emission source is negligible.  
Extinction of H$_2$ emission could occur within the disk if dust is mixed
with the warm gas,
although \citet{Cal02} found that small
dust grains are underabundant within 5 AU of the star.
We assume an interstellar
extinction law from \citet{Car89}, for grains typical of the interstellar
medium ($R_V=3.09$).

For each H$_2$ progression, the average depth $d_{abs}$ of the initial \Lya\
absorption depends on the opacity of the lower level of H$_2$ (see Fig. 17
and Table 2).  If a \Lya\ photon is, on average, absorbed close to the disk
surface, then the H$_2$ lines excited by that photon will not be significantly
attenuated by any disk extinction.  Thus, the extinction varies for the
different progressions.
In Figure 16,
$A_V$ represents the extinction through half of the slab (not the extinction
through the entire slab or the extinction from the star to the observer), 
which is equal to the largest
possible average extinction of H$_2$ emission when \Lya\ photons are
absorbed uniformly throughout the slab.
Disk extinction attenuates both the incident \Lya\ emission
and the outgoing H$_2$ emission, but not the observed \Lya\ emission, and thus increases the already large 
filling factor.  
In the context of our model, significant disk 
extinction can be ruled out because the computed  
filling factors would be much larger than unity.

Interstellar extinction affects the fluxes smoothly across the wavelength
region, and 
does not significantly impact the derived filling factor because it
attenuates both the observed \Lya\ emission and the H$_2$ emission.  If we consider only extinction and ignore
opacity in the H$_2$ lines, then from fitting H$_2$ line fluxes we obtain a
minimum $\chi^2_{\nu,H2}=6.06$ at $A_V$=1.3, which is significantly larger than the
$\chi^2_{\nu,H2}$ for the best-fitting models.

\subsection{Adoption of the Thick Disk Geometry}
Based on these results, and the results presented in Figs. 15-16 and Table
5, we adopt the thick disk geometry (Geometry 2 in Table 5) as the
preferred geometry to describe the H$_2$ emission.  The thick disk geometry
is physically plausible and produces reasonably good fits with self-consistent results, although
certain other geometries can produce lower values of $\chi^2$ in fits to H$_2$ fluxes or
in the \Lya\ fits.  Some of those geometries (e.g., Geomteries 7 and 8 from
Table 5) represent morphologies that
are difficult to reconcile with our previous conclusion that the H$_2$
emission occurs in a disk.  In other geometries (e.g., Geometries 1 and 11 from Table 5),
contours from the H$_2$ flux fitting do not overlap with the contours from
the \Lya\ fitting as well as they do for the thick disk geometry, i.e. the
joint probability is lower (see Fig. 11).

We caution the reader that the thick disk geometry is best only in the
context of our restricted set of models.  Warm H$_2$ in the TW Hya
star-disk system could in principle have a morphology not considered here.

%%%%%%%%%%%%%%%%%%%%%%%%%%%%%%%%%%%%%%%%%%%%%%%%%%%%%%%%%%%%%%%%%%%%%%%

\section{DISCUSSION}
\subsection{Synthesizing the H$_2$ and Ly$\alpha$ modelling results}
Figures 11 and 13, and Table 5, compare results from our analysis of
individual H$_2$ line
fluxes with our results from the \Lya\ profile reconstruction.
  Based on the overlap
region of acceptable models for both the individual H$_2$ line fluxes and
the reconstructed \Lya\ profile,
we conclude that
a column density of $\log N($H$_2)=18.5^{+1.2}_{-0.8}$ is heated to a
temperature of $2500^{+700}_{-500}$~K, with a filling factor of $\eta=0.25\pm0.08$
around TW Hya ($2\sigma$ error bars).
Table 2 lists the percentage $P_{dis}$ of H$_2$ that
dissociates from the upper level, the H$_2$ dissociation rate in terms of
mass ($M_{dis}$),
and the total flux for each
progression ($F_{mod}$) for our best-fit model of $\log N($H$_2)=18.5$ and
$T=2500$~K.  The notion of a single temperature is a simplistic assumption,
but it is sufficient to reproduce our data.  The temperature derived here is
warmer than the $\sim1100-1300$ K temperatures seen in the fundamental CO
emission by \citet{Naj03}.

The mass column corresponding to
 $\log N($H$_2$)=18.5 is $10^{-5}$ g cm$^{-2}$,  
which is about $10^{-7}$ times smaller than the mass column 
predicted to be within 1 AU by \citet{Dal99}. 
This suggests that H$_2$ fluorescence probably occurs in a very 
shallow surface layer of the disk, which could be a warm, extended disk
atmosphere.  The \Lya\ fluorescence 
process cannot be used to detect
cold H$_2$, which most likely represents the bulk of the mass in the disk.

In our models, we assume that H$_2$ gas is thermalized, so that relative
level populations depend only on temperature.  This is reasonable because 
the gas
density at the disk surface should be between $n_H=10^6-10^8$~cm$^{-3}$
\citep{Gom00}, which is far larger than 
the critical density of $n_H$=$10^{4}$~cm$^{-3}$ at which 
collisions typically dominate over quadrupole radiative
de-excitation.  Below the critical density,
collisions cannot repopulate excited H$_2$ levels as fast
as they radiatively deexcite. Fluorescent progressions
pumped out of higher energy states would be weaker than
expected from our thermal models, generating scatter
that we do not see in our reconstructed \Lya\ profile.
This conclusion confirms the assumption of thermally 
excited H$_2$ gas used by \citet{Bla87} and \citet{Bur90} to model H$_2$ fluorescence.

\subsubsection{Filling factor of H$_2$ around the \Lya\ source}
Figure 13 shows that the filling factor of the H$_2$ around the \Lya\ emitting
region is $\eta=0.25\pm0.08$.  This factor
represents the solid angle of the sky subtended by H$_2$ as seen from the
source of \Lya\ photons near TW Hya,
although other geometrical effects may contribute.  Neutral hydrogen mixed with the
warm H$_2$ could scatter \Lya\ photons, increasing the effective path
length of the \Lya\ photons and thereby reducing the calculated filling
factor.  Any directional dependence of the  \Lya\ emission, and the percentage
 of \Lya\ emission visible to us versus the percentage visible to the
 warm H$_2$, will also affect the calculated filling factor.
A total of 1-2\% of the intrinsic
\Lya\ emission is reprocessed into H$_2$ emission (see Table 6 and \S4),
which places a lower limit of 0.01 on the filling factor.

For any flared disk geometry, the filling factor can be converted into a
disk height by comparing the surface areas of a cylinder and a sphere.
For $\eta=0.25$, the implied geometrical disk height from the
midplane to the surface at distance $R$ from the center of the star is 
H=$0.25 (\frac{R}{1 AU})$ AU, compared with predicted disk heights from
\citep{Dal01} of about 0.15~AU at a distance of 1~AU from the star.

\subsubsection{Relationship between UV and IR H$_2$ Emission}
Emission in the 1-0~S(1) rovibrational transition at 2.1218 $\mu$m has been 
detected from 4 T Tauri stars \citep{Bar03}.  The narrow width and absence
of any velocity shift with respect to the stars suggests a disk origin of
this emission.  The flux in this line from an optically thin emission region
\begin{equation}
F=\frac{N(1,3) \pi r^2 A_{ul}}{4\pi d^2}\frac{hc}{\lambda},
\end{equation}
where $N(1,3)$ is the H$_2$ column density of the disk in the
$\Vlo=1,\Jlo=3$ level, $\pi r^2$ is the projected surface area of the
emission region, and $A_{ul}=3.47 \times
10^{-7}$ s$^{-1}$.  If the 1-0 S(1) line flux of $1\times10^{-15}$ erg
cm$^{-2}$ s$^{-1}$
measured by Weintraub et al. (2000) comes from a heated disk
surface layer with N(H2)=18.5 and T=2500 K, then the 1-0 S(1)
emission must extend out to at least 5 AU for a face-on disk.  
This corresponds to an angular extent of
$\pm0\farcs1$, which could have been resolved in our HST
observations, but was not. Thus, it seems likely that at
least some of the 1-0 S(1) emission comes from H$_2$ gas
that is not fluoresced by \Lya.
Some of the emission in the 1-0 S(1) line may be
produced deeper in the disk or farther from the star, as a result of X-ray
ionization and
subsequent collisional excitation of H$_2$ by non-thermal electrons
\citep{Wei00}.  

\subsubsection{H$_2$ Dissociation Rate}
From the flux in observed transitions, we calculate a dissociation rate of 
$9.8\times 10^{-11}$ $M_\odot$~yr$^{-1}$ for H$_2$ in the disk.
Forward modelling of undetected Lyman and Werner-band transitions 
excited by \Lya\ photons increases the calculated dissociation 
rate to $1.6\times 10^{-10}$ $M_\odot$ yr$^{-1}$.
Depending on the timescale for collisional de-excitation of H$_2$, 
the actual dissociation rate could be much 
higher due to multiple pumping [pumping from lower rovibrational levels of
H$_2$ that are populated by the initial pump and subsequent fluorescence -- see \citet{Shu78} for details], which can populate highly excited 
rovibrational levels in the B electronic state that have dissociation
probabilities as high as 50\%.

The H$_2$ dissociation rate we estimate is within a factor of 10 of TW
Hya's mass accretion rate (see \S 2.1), meaning that the dissociation
provided by \Lya\ fluorescence may be an important process in the accretion
of hydrogen onto the star.  The neutral H released by dissociation will
cause the
incident \Lya\ to scatter, increasing the probability of absorption in the
pumping transitions and therefore decreasing the required filling factor.
The  neutral H could also be
sufficiently optically 
thick to produce the self-reversal of the \Lya\ profile
that irradiates the disk (see Fig. 14).

%%%%%%%%%%%%%%%%%%%%%%%%%%%%%%%%%%%%%%%%%%%%%%%%%%%%%%%%%%%%%%%

\subsection{FUV radiation field}
The FUV radiation field incident on the accretion disk can significantly
affect the chemistry and temperature structure of the disk.  In 
Table 6 we list the flux for spectral features in the FUV spectrum of TW Hya.
The observed \Lya\ line contributes 67\% of the observed FUV flux of
$6.5\times10^{-11}$~ergs~cm$^{-2}$~s$^{-1}$ between 1170--1700 \AA.
From the reconstructed \Lya\ profile, we infer an
intrinsic flux
of $8-16\times10^{-11}$~ergs~cm$^{-2}$~s$^{-1}$, depending on the
correction for the central absorption.  With this correction, \Lya\
emission contributes 80--90\% of the FUV flux of TW Hya, of which 1-2\% is
reprocessed into H$_2$ emission.
The continuum accounts for
about $6\times10^{-12}$~ergs~cm$^{-2}$~s$^{-1}$, or about 4\% of the total
emission, including the estimated intrinsic \Lya\ flux.  The remaining flux
occurs in other emission lines such as the \ion{C}{4} 1550~\AA\ doublet.
The UV flux below 1170~\AA\
measured by \FUSE\ adds
at most 2\% additional flux to the strength of the radiation field.  The UV
radiation
between 1700--2000~\AA\ may
add to the radiation field, primarily due to emission in lines such as
\ion{Si}{2}]~$\lambda$1817 and \ion{C}{3}]~$\lambda$1909, but will have
little effect on the excitation of molecules and atoms in the disk. 
The estimated total emission after including  \Lya\
flux that is not observable is approximately $10^{-3}$
$L_\odot$, assuming  isotropic emission and a distance of 56 pc.

The properties of the 
disk, both in the PDR-like disk atmosphere and in the cooler outer regions, 
will be partially controlled by the \Lya\ emission, as evidenced by 
the fluorescent H$_2$ emission.  Not only does \Lya\ dominate the FUV flux
of TW Hya, but it can also scatter off neutral H towards the interior of
the disk, allowing it to penetrate the disk further than continuum UV
radiation. \Lya\ can dissociate molecules such as H$_2$ and H$_2$O, and can ionize Si
and C.  \citet{Ber03} explain that the enhancement of 
CN relative to HCN occurs in disks because Ly$\alpha$ radiation can
photodissociate HCN, whereas CN can be dissociated only by radiation with $\lambda<1150$ \AA.

\subsection{Heating mechanisms}
We have found evidence for a warm surface layer of H$_2$ 
on the disk around TW Hya.
From the lack of spatial extent in the \STIS\ echelle 
cross-dispersion direction, the
heated surface is located within 2 AU of the star.
Models of dust in disks of CTTSs typically predict $T\approx 1000$ K only within
0.1 AU of the star \citep{Dal01a,Chi97}.  The FUV
radiation field at 1 AU from the star is approximately $10^7$ $G_0$, where
the Habing field $G_0=1.6\times 10^{-3}$ erg cm$^{-2}$ s$^{-1}$ (the
local interstellar field is 1.7 $G_0$), which is stronger than the FUV
radiation field incident upon most PDRs ($10^3-10^6$ $G_0$).  \citet*{Hol00}
predicted from models of photo-dissociation regions
that a warm atmosphere-like structure, heated by UV radiation,
should surround the inner
regions of a disk, with a temperature between 500 and 3000 K.  

Because of the absence of micron-sized dust grains within 5 AU of TW Hya
\citep{Cal02}, a large percentage of the FUV radiation may be deposited
into the gas.
In their analysis of the H$_2$ fluorescence from Mira B, \citet{Woo02} found that the 
temperature increases sharply with radius from the UV source,
 and that the \Lya\ fluorescence process 
is a significant heating source for the gas.  The energy deposited into the 
warm gas near TW Hya due to Ly$\alpha$ photoexcitation and subsequent 
fluorescence is  $1.4\times 10^{29}$ erg s$^{-1}$, which corresponds 
to roughly 0.01 eV s$^{-1}$ per molecule of H$_2$.
The strong X-ray luminosity of TW Hya, $L_X=3.6\times10^{-4}~L_\odot$ 
between 0.45--6.0~keV \citep{Kas02}, could heat the disk surfaces by an
additional 500 
K \citep{Ige99,Gla01,Fro02} by ionizing atoms to produce a large reservoir
of energetic non-thermal electrons \citep[e.g.,][]{Mal96,Yan97}.
X-rays can also penetrate
deeper into the gas than UV radiation and, consequently, may heat gas
 deeper in the
disk than the UV radiation.

Certain observed fluorescence progressions are pumped from 
lower rovibrational levels with large energies, such as $\Vlo=3, \Jlo=25$
($E^{\prime\prime}=4.2$ eV)
and $\Vlo=5, \Jlo=18$ ($E^{\prime\prime}=3.8$ eV).  The measured
flux in these progressions, comibed with the observed flux at the pumping wavelengths, requires that of order 1\% of the H$_2$
resides in these levels.  This population is too large to be
thermally populated
at any reasonable temperature for the H$_2$ gas.  Therefore, non-thermal
processes must be populating these levels.
Although the fluorescence process can populate
high vibrational levels in the ground electronic state,
fluorescence cannot directly populate high rotational levels, and no direct 
paths exist to the levels  $\Vlo=3, \Jlo=25$ and $\Vlo=5, \Jlo=18$.
H$_2$ can form in highly excited states \citep*{Kok01,Str01} via
dissociative recombination of H$_3^+$.  Given a dense medium and a large
ionizing flux,
then H$_2^+$ may be present in large enough quantities to react with H$_2$
to produce H$_3^+$ in highly excited states.  More attempts to detect H$_3^+$ would be useful.

%%%%%%%%%%%%%%%%%%%%%%%%%%%%%%%%%%%%%%%%%%%%%%%%%%%%%%%%%%%%%%%

\section{SUMMARY}
\label{section:summary}

We have modeled a plane-parallel slab of warm H$_2$ (see Fig. 1), which
represents the surface of a protoplanetary disk irradiated by UV flux from
the star.
Our analysis of the ultraviolet spectrum of TW Hya obtained with the \STIS\
instrument on \HST\ and with \FUSE\ leads to the following conclusions:

1.  The FUV continuum rises shortward of 1700 \AA, which is indicative of
an H$_2$ dissociation continuum, although it could also be produced by
H$_2$ fluorescence due to FUV continuum pumping or an
additional accretion or activity component.  The FUV continuum is not well explained by
simple models of a pure hydrogen slab, which are commonly invoked to analyze the excess
NUV continuum.

2.  The extinction towards TW Hya is negligible, based on the hydrogen
column density in our line of sight and assuming an interstellar gas-to-dust ratio.

3.  Self-absorption of Lyman-band transitions involving low excitation energy levels weakens the flux in these lines.
We model this effect by simulating \Lya\ emission
entering a plane-parallel atmosphere and pumping the H$_2$.

4.  Using the observed H$_2$ fluxes and our 
fluorescence models, we reconstruct the
\Lya\ line profile incident upon the warm molecular layer and compare it to
the observed \Lya\ profile, for a range of assumed temperatures and column
densities.  
Undetected
progressions rule out a large region of parameter space in this model.  The
reconstructed \Lya\ profile is similar to the observed profile in the
wings, but shows a much narrower absorption feature than is observed.
This narrow absorption component in the reconstructed profile could be  
a self-reversal or a component of the wind between the
source of \Lya\ emission and the warm molecular region.

5.  Our models indicate that a molecular layer with a kinetic temperature 
of $2500^{+700}_{-500}$ K and a column density
of $\log N$(H$_2)=18.5^{+1.2}_{-0.8}$ ($2\sigma$ error bars) 
absorbs \Lya\ radiation in the surface layer
and inner edge of the disk within 2 AU of the central star.  
The \Lya\ pumping leads to a small H$_2$ dissociation rate and  does not cause 
significant disk dissipation.
The filling factor of the warm H$_2$ around TW Hya is $\eta=0.25\pm0.08$,
although significant
uncertainties in the geometry of the fluorescent H$_2$ weaken our confidence in the large filling factor.

6.  The warm H$_2$ most likely resides in a warm surface layer of the disk.  
This surface layer may be analgous to a classic PDR, although in this case the FUV radiation field is dominated by emission lines rather than a continuum.  In
particular, \Lya\ comprises about 85\% of the FUV radiation field below
2000 \AA, and it controls the excitation and ionization of the disk
surface.

7.  Some of the observed H$_2$ lines are pumped from high rotational levels
that cannot be excited thermally or by fluorescence. 
Formation of rotationally excited H$_2$ by reactions with
H$_3^+$ may explain these emission lines.

%%%%%%%%%%%%%%%%%%%%%%%%%%%%%%%%%%%%%%%%%%%%%%%%%%%%%%%%%%%%%%%

\acknowledgements

This research is supported by NASA grant S-56500-D to the University of
Colorado and NIST, and by a grant to the University of Colorado of NASA
funds through the Johns Hopkins University.  We thank the referee for
thorough and valuable comments.  The referee especially helped to
significantly improve
the clarity of the paper.
We also thank 
Chris Greene and Ben McCall for valuable
discussions about H$_2$ formation, and
Phil Maloney for insightful comments about PDRs.

%%%%%%%%%%%%%%%%%%%%%%%%%%%%%%%%%%%%%%%%%%%%%%%%%%%%%%%%%%%%%%%

% Bibliography

\begin{figure}
\plotone{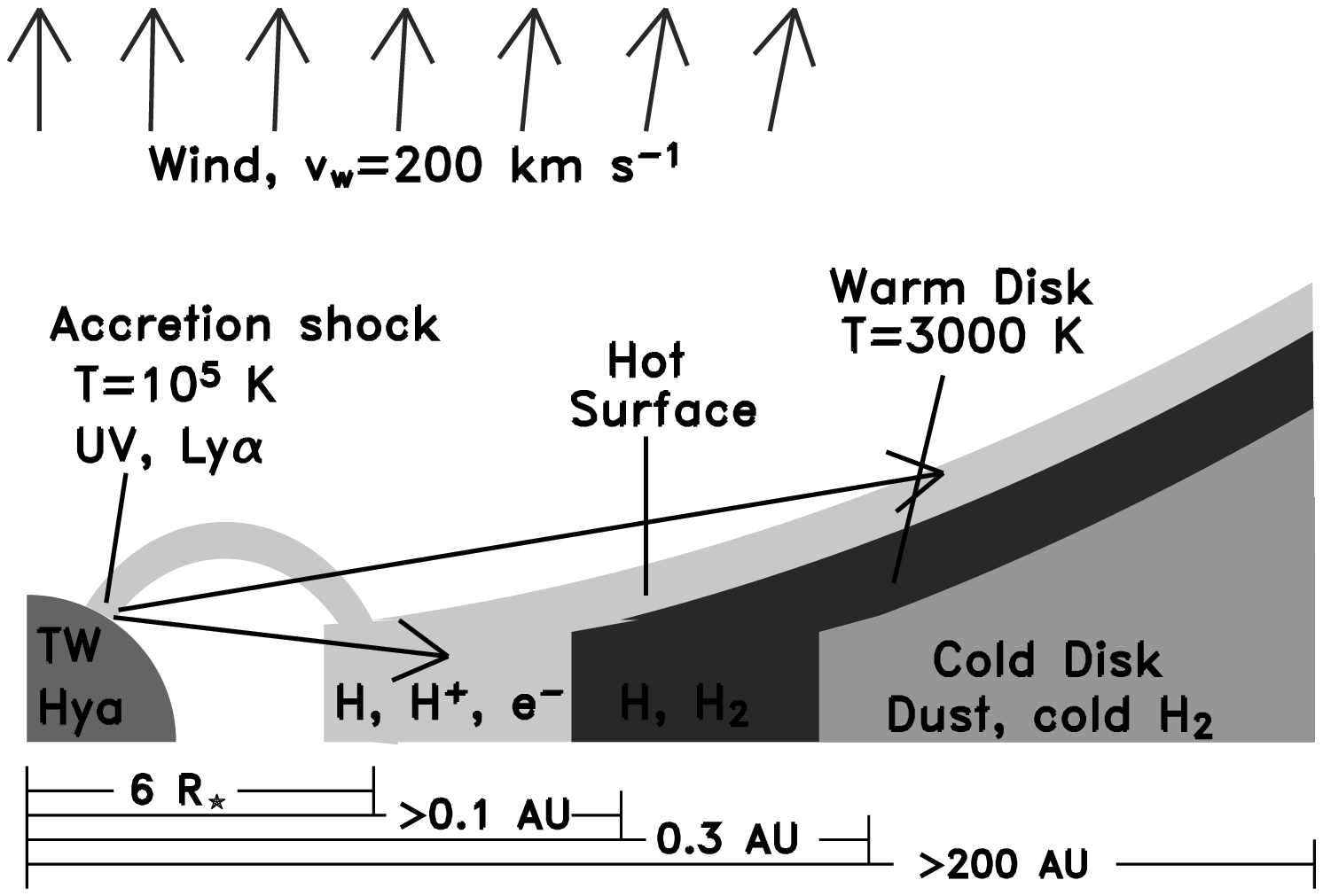}
\label{fig:twfig.ps}
\caption{A schematic diagram of our model of
TW Hya and its disk (not to scale).
FUV radiation, primarily \Lya, is produced at the accretion shock and heats
the disk.  The warm H$_2$ is located in surface layers of the disk within 2
AU of the star.
The hot surface of the disk, composed of atoms and ions, probably extends
inwards to about $6 R_*$ \citep{Joh01a}, where it is truncated by the
stellar magnetosphere.}
\end{figure}

\begin{figure}
\plotone{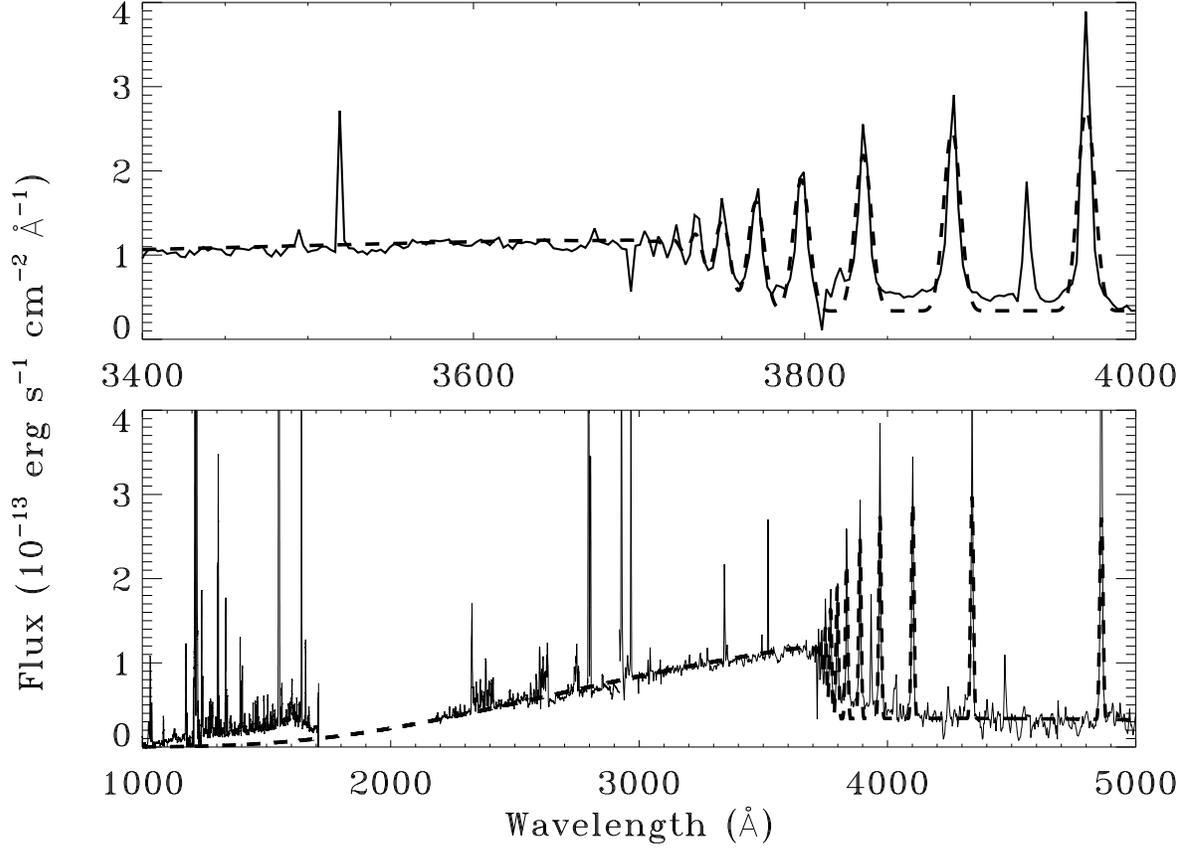}
\label{fig:accretion.eps}
\caption{The solid line shows the observed spectrum of of TW Hya, after
subtracting the template spectrum of V819 Tau. 
The 
dashed line shows the synthetic spectrum of a pure hydrogen slab fit to the
observed Balmer jump and the Balmer lines.
The pure hydrogen continuum accurately approximates
 the NUV continuum of TW Hya 
but significantly underestimates the FUV continuum.  The rise of the FUV
continuum shortward of 1700 \AA\ is indicative of an H$_2$ dissociation
continuum produced by collisions with energetic electrons, although it
could also be produced by H$_2$ fluorescence due to FUV pumping or 
an additional accretion or activity component.}
\end{figure}

\begin{figure}
\resizebox{6.in}{6.in}{\plotone{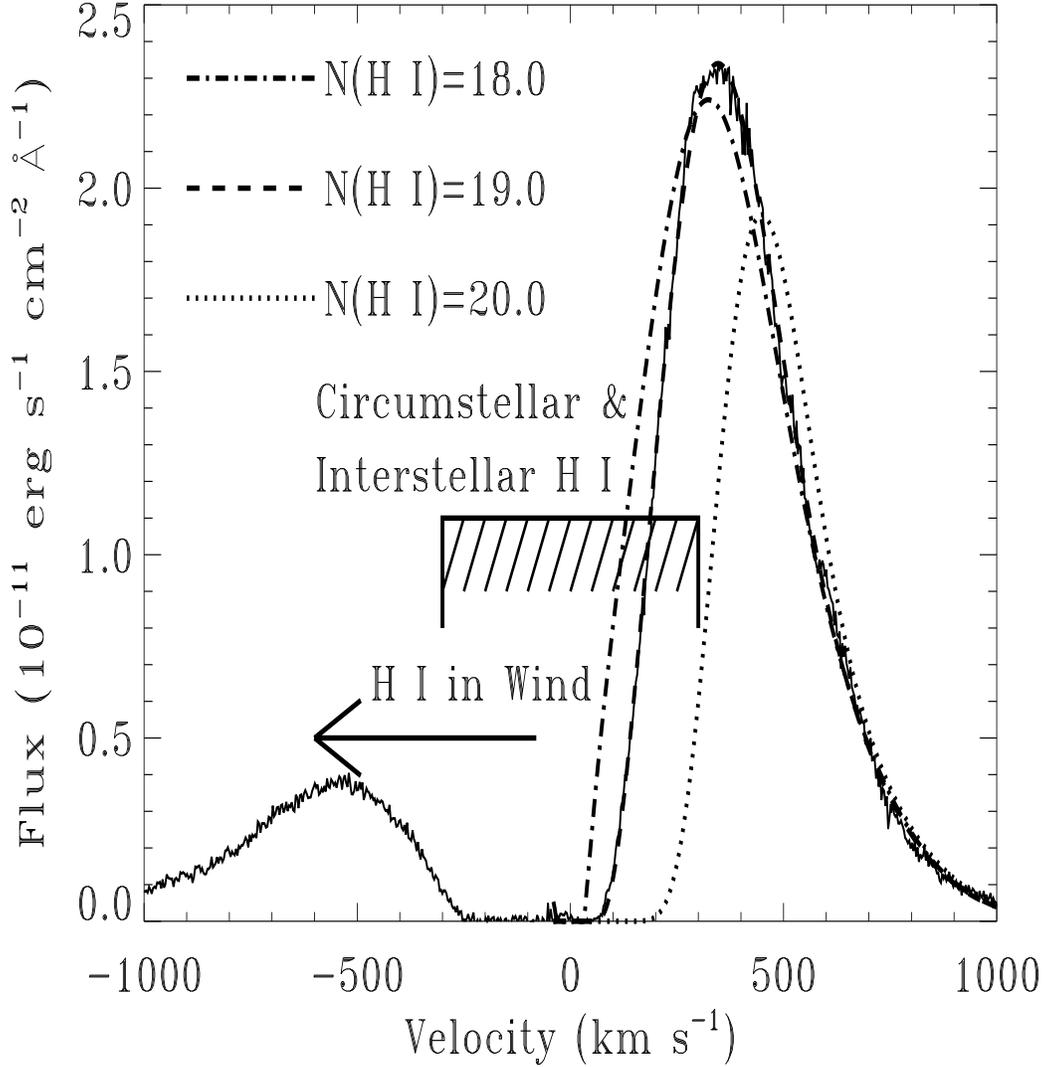}}
\label{fig:lya_velocity.eps}
\caption{The \Lya\ emission from TW Hya is partially
 absorbed by \ion{H}{1} in the
line of sight to TW Hya. The interstellar and circumstellar media, centered 
at about 0 \kms, account for the red side of the absorption.  The blue
side of the absporption is produced by \ion{H}{1} in the wind.
The dashed line shows
our best fit to the observed \Lya\ profile (solid line)
using two Gaussian emission profiles to model the intrinsic \Lya\ line before
the absorption, while the dotted and dashed-dotted lines assume different
\ion{H}{1} column densities that poorly fit the red
 side of the observed profile.}
\end{figure}

\begin{figure}
\resizebox{5.in}{5.in}{\plotone{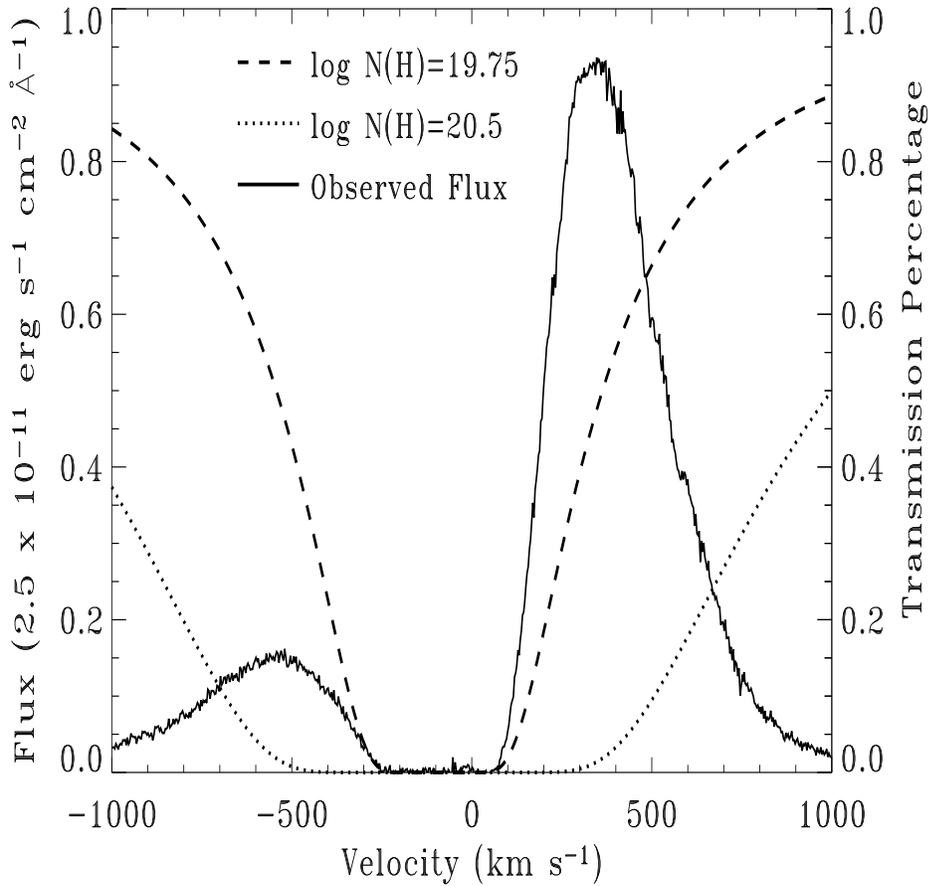}}
\label{fig:o6_h2abs.eps}
\caption{The transmission percentage across the observed \Lya\
profile (solid) for column densities
$\log N$(H)$=19.75$ (dashed) and $\log N$(H)$=20.5$ (dotted), assuming no Doppler
broadening.  This assumption places a strong upper limit of $N($\ion{H}{1}$<19.75$) 
towards TW Hya.}
\end{figure}

\begin{figure}
\resizebox{5.in}{5.in}{\plotone{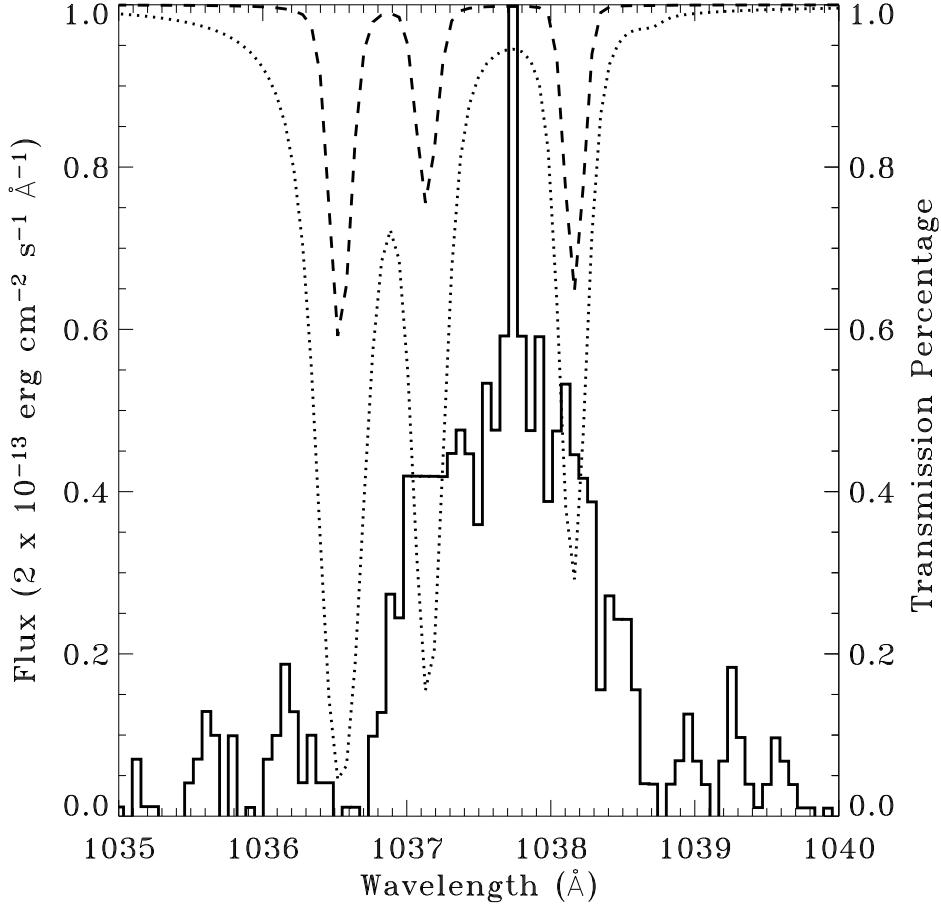}}
\label{fig:o6_h2abs.eps}
\caption{The O VI emission line from TW Hya (solid), as observed with the LiF1a
detector on \FUSE.  
We place a limit of $N($H$_2$)$<18$ in our line of sight towards
TW Hya by comparing the observed emission with the simulated 
transmission percentage
given the resolution of \FUSE, 
through a medium of H$_2$ with $T=100$~K and $N($H$_2)$=17.5
(dashed) and $N($H$_2)$=19.0 (dotted).
We would observe many
H$_2$ absorption lines against the O VI emission from TW Hya
 for hotter excitation
temperatures of H$_2$.}
\end{figure}

\begin{figure}
\resizebox{5.in}{5.in}{\plotone{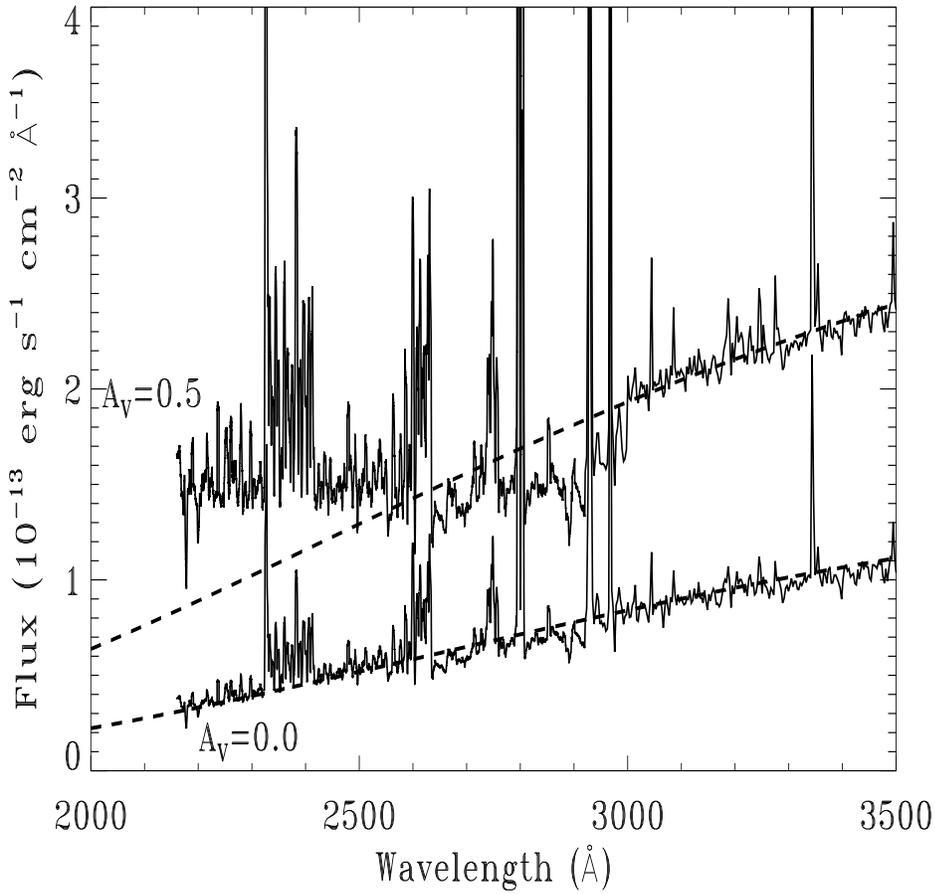}}
\label{fig:o6_h2abs.eps}
\caption{The NUV continuum (solid) is well fit by the accretion model (dashed) for
$A_V=0.0$, but if $A_V=0.5$, the rise in the continuum below 2600 \AA\ is
not fit well.
}
\end{figure}

\begin{figure}
\resizebox{6.in}{6.in}{\plotone{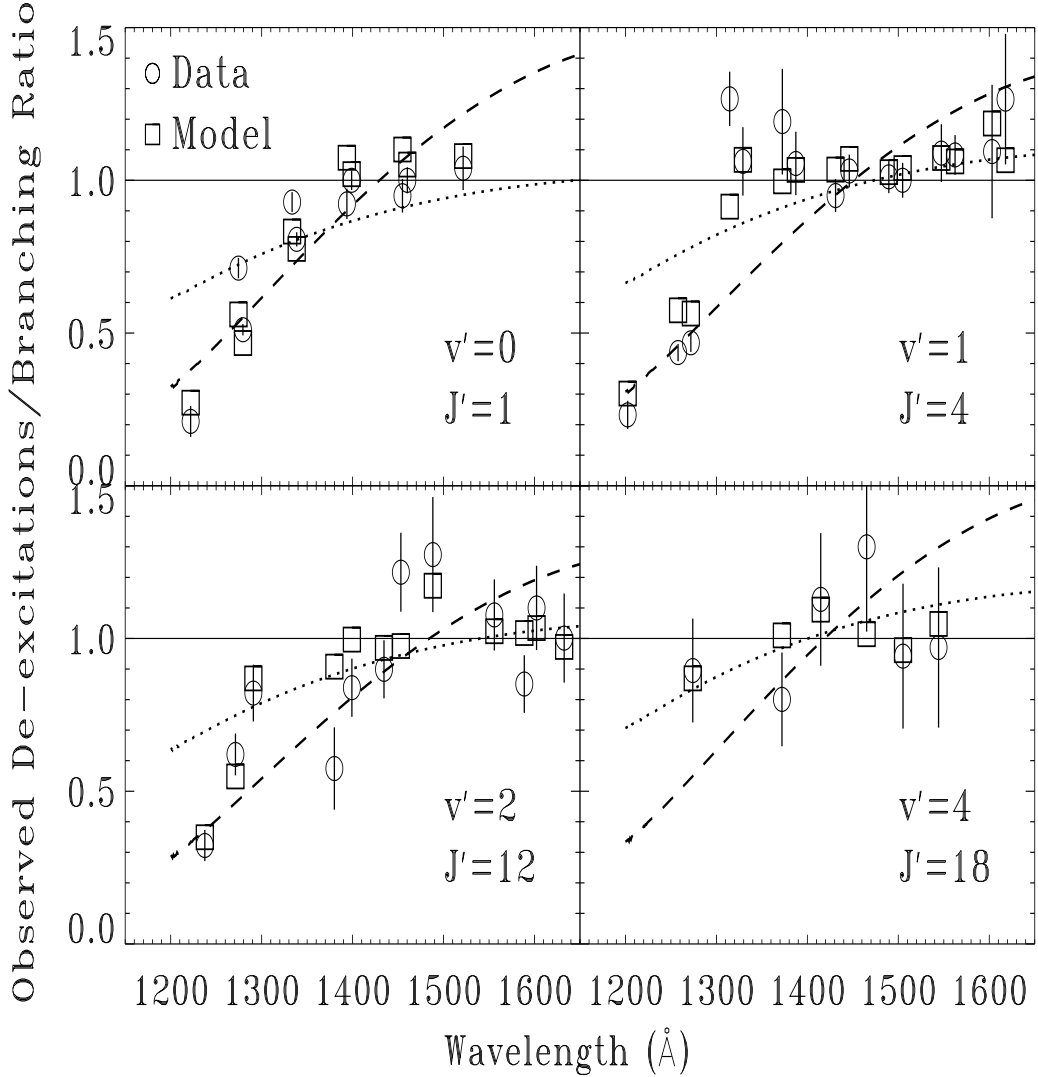}}
\label{fig:extinct.ps}
\caption{For four different upper levels, we plot the normalized ratio of the observed 
de-excitations in each H$_2$ line divided by the respective branching ratio 
(circles).  The squares show the fluxes prediced based on the optical
depth in each lower level 
for our adopted model for $\log N($H$_2)=18.5$
and $T=2500$~K (see \S5).  
The dotted 
and dashed lines show extinction curves for $A_V=0.5$ and $1.5$, 
respectively. 
Emission in an
optically thin medium, with no intervening extinction, would result in
ratios of 1.0.  The apparent wavelength dependence results 
from the larger optical depth in those levels of H$_2$ which have lower excitation energies.
The H$_2$ line fluxes from \Vup=4, \Jup=18 show
no wavelength dependence because the observed lines all have lower levels
with energies large enough that they are not populated.}
\end{figure}

\pagebreak

\begin{figure}
\resizebox{6in}{4in}{\plotone{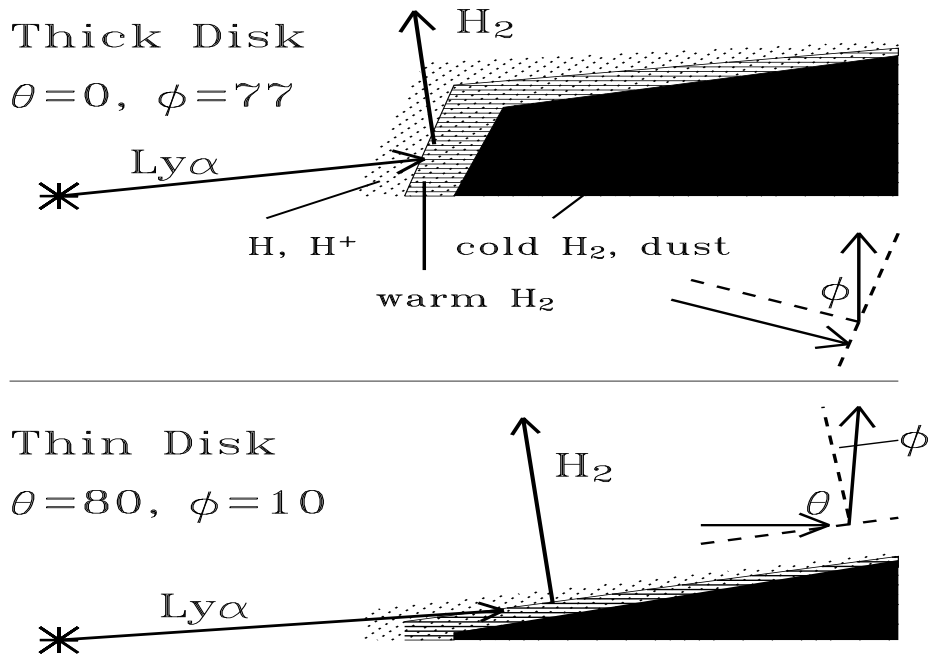}}
\label{fig:disks_mods.eps}
\caption{Two plausible geometries for H$_2$ fluorescence in the surface
layers of
a protoplanetary disk.  The \Lya\ enters the surface layer of H$_2$ at angle $\theta$ 
with respect to the normal and is reprocessed into H$_2$ emission, which escapes the disk at angle $\phi$.}
\end{figure}

\begin{figure}
\resizebox{5.5in}{4in}{\plotone{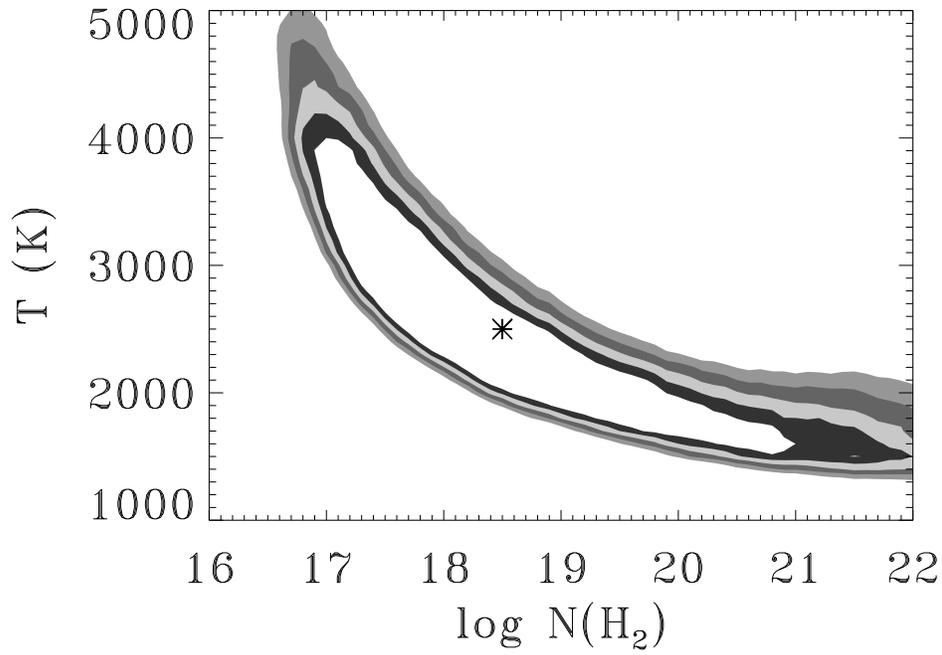}}
\label{fig:plotback.eps}
\caption{$1-5\sigma$ confidence contours calculated from $\Delta \chi^2_{\nu,H2}$
 for fits to the individual H$_2$ line fluxes for the thick disk model (see \S 3).  The
star indicates the position of the model with $T=2500$ K and $\log
N$(H$_2$)=18.5 (see \S6).}
\end{figure}

\begin{figure}
\resizebox{4in}{6in}{\plotone{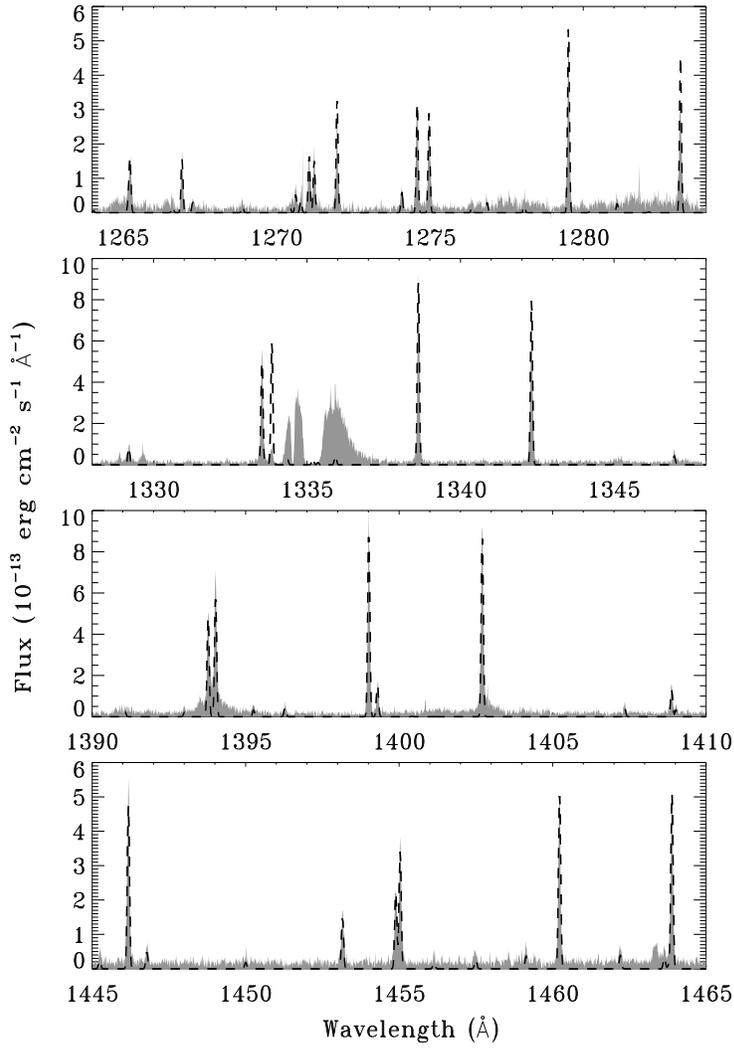}}
\label{fig:make_h2flxs.eps}
\caption{Selected regions of the \HST/\STIS\ spectrum of TW Hya.  The narrow features 
are all H$_2$ lines.  The dashed lines show our model fit to the H$_2$ emission.  The 
line at 1333.8 \AA\ is supressed by \ion{C}{2} in the wind of TW Hya, which 
places the H$_2$ interior to the wind.}
\end{figure}

\clearpage

\begin{figure}
\resizebox{6.5in}{6.5in}{\plotone{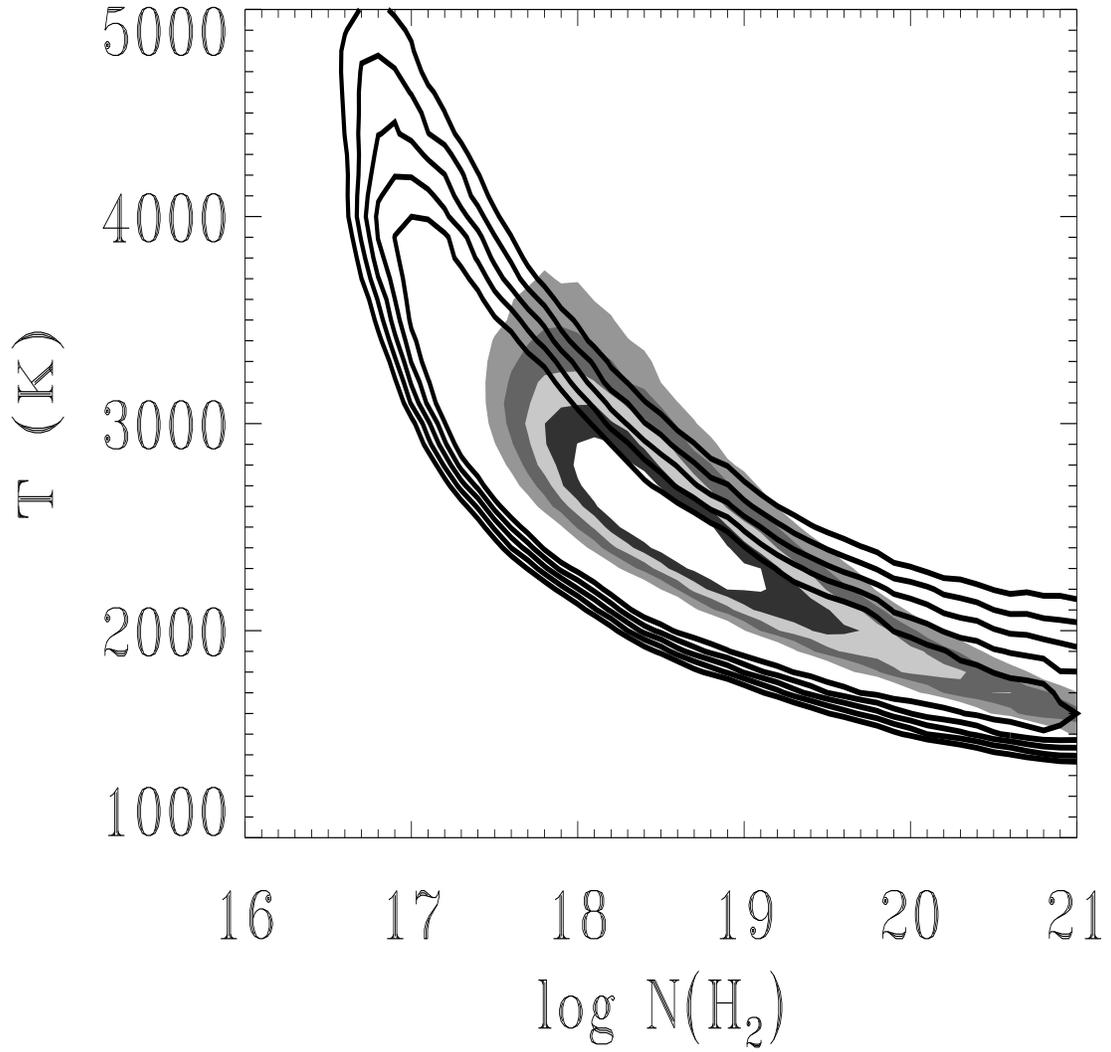}}
\label{fig:lya_model1.eps}
\caption{$1-5\sigma$ confidence contours calculate from $\Delta\chi^2_{\nu,Ly\alpha}$ (shaded regions)
for our fit of the reconstructed
\Lya\ profile to the red wing of the observed \Lya\ profile, compared with
$1-5\sigma$ confidence contours (solid lines) for fits to the observed H$_2$ fluxes.}
\end{figure}

\begin{figure}
\resizebox{6.5in}{6.5in}{\plotone{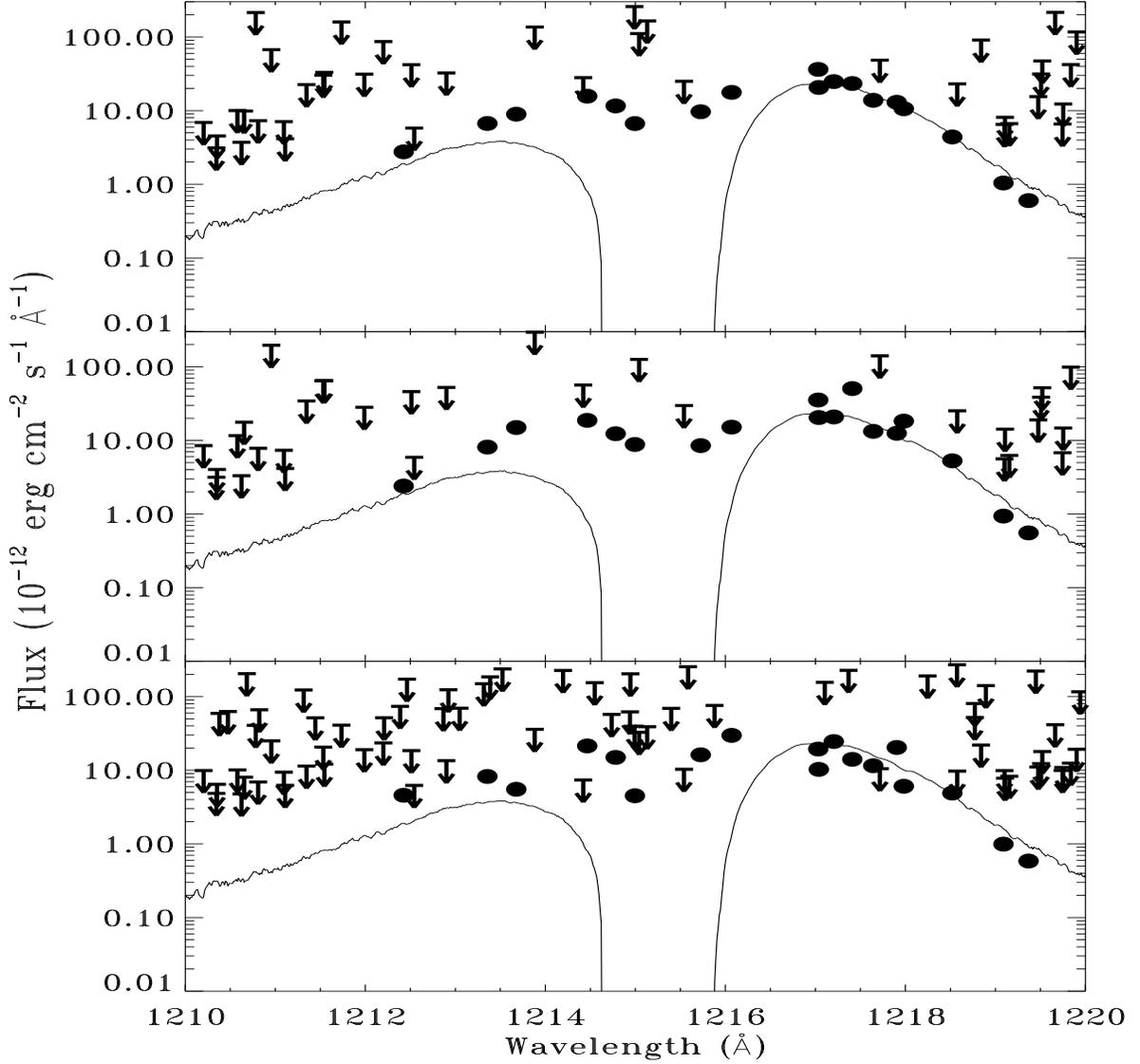}}
\label{fig:lya_pap2_final.eps}
\caption{Comparison of the observed (solid line) and calculated (circles) 
\Lya\
emission profiles, for models with (top) $\log N($H$_2$)=18.5, $T=2500$ K,
and $\eta=0.25$
(middle) $\log
N($H$_2$)=19.0 $T=2000$ K, and $\eta=0.41$; and (bottom)
 $\log
N($H$_2$)=19.0 $T=3000$ K, and $\eta=0.20$.
The arrows show upper limits to the calculated
\Lya\ emission based on undetected H$_2$ progressions.  The
top plot is the best fit because of the least scatter in the red
wing of \Lya.  The bottom model is unacceptable
because several undetected progressions ($\Vup=7,\Jup=4$ pumped at 1214.4 \AA\ and
$\Vup=5,\Jup=20$ pumped at 1217.7 \AA) imply upper limits of \Lya\ flux that are
significantly lower than the flux reconstructed from nearby detected
progressions.}
\end{figure}

\begin{figure}
\resizebox{6.5in}{6.5in}{\plotone{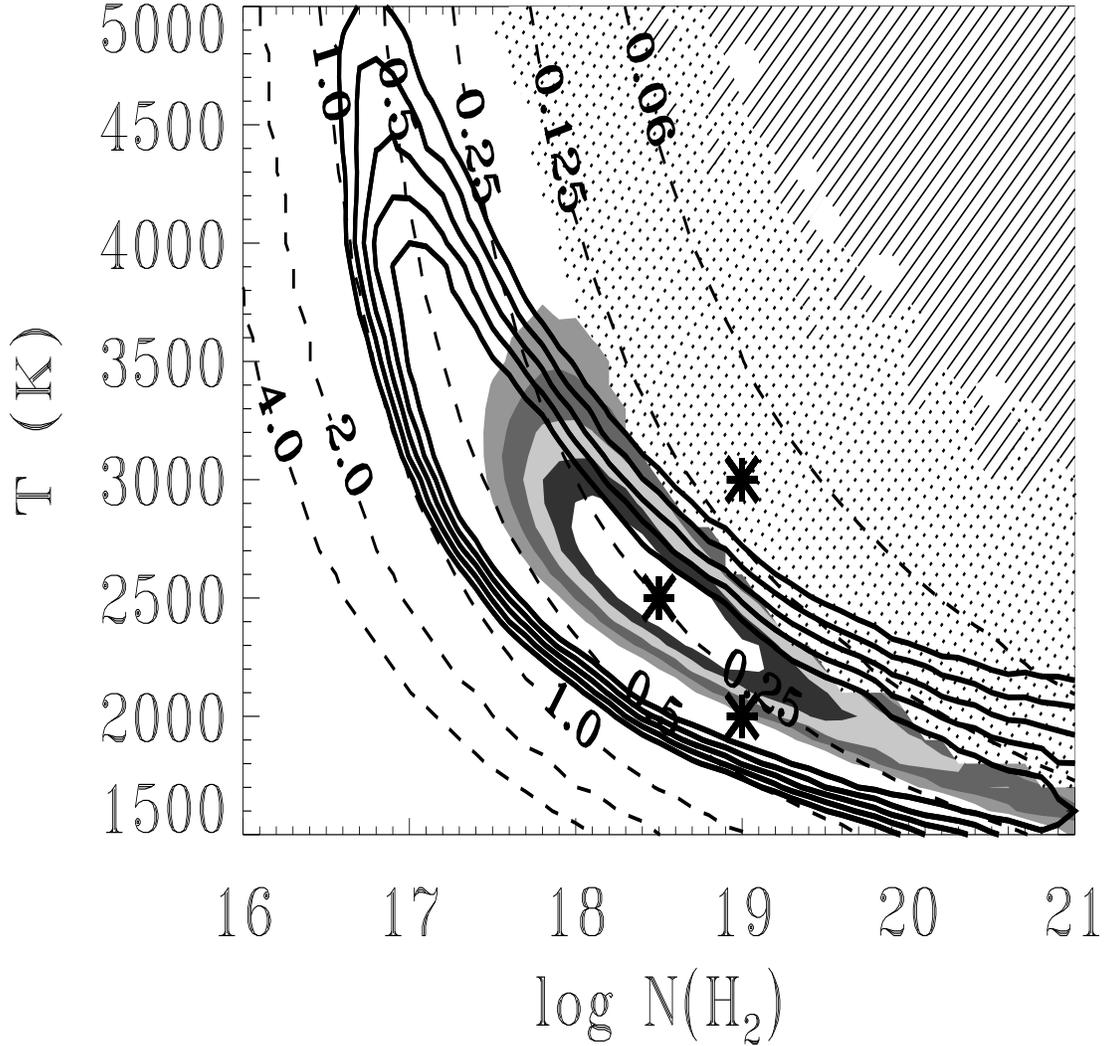}}
\label{fig:contour_total_final.eps}
\caption{Similar contours as in Figure 11.  The dots indicate the parameter 
space that we rule out because those parameters predict that lines pumped by 7-4 P(5)
should be observed, and the diagonal
lines indicate the
parameter space ruled out because many additional other progressions should be
observed but are not detected.
Filling factor contours (dashed lines) of 0.125--4.0 are also shown.  A filling factor
above 1 is unphysical.  The three models with
reconstructed \Lya\ profiles in Fig. 13 are shown as asterisks.}
\end{figure}

\begin{figure}
\resizebox{6.5in}{6.5in}{\plotone{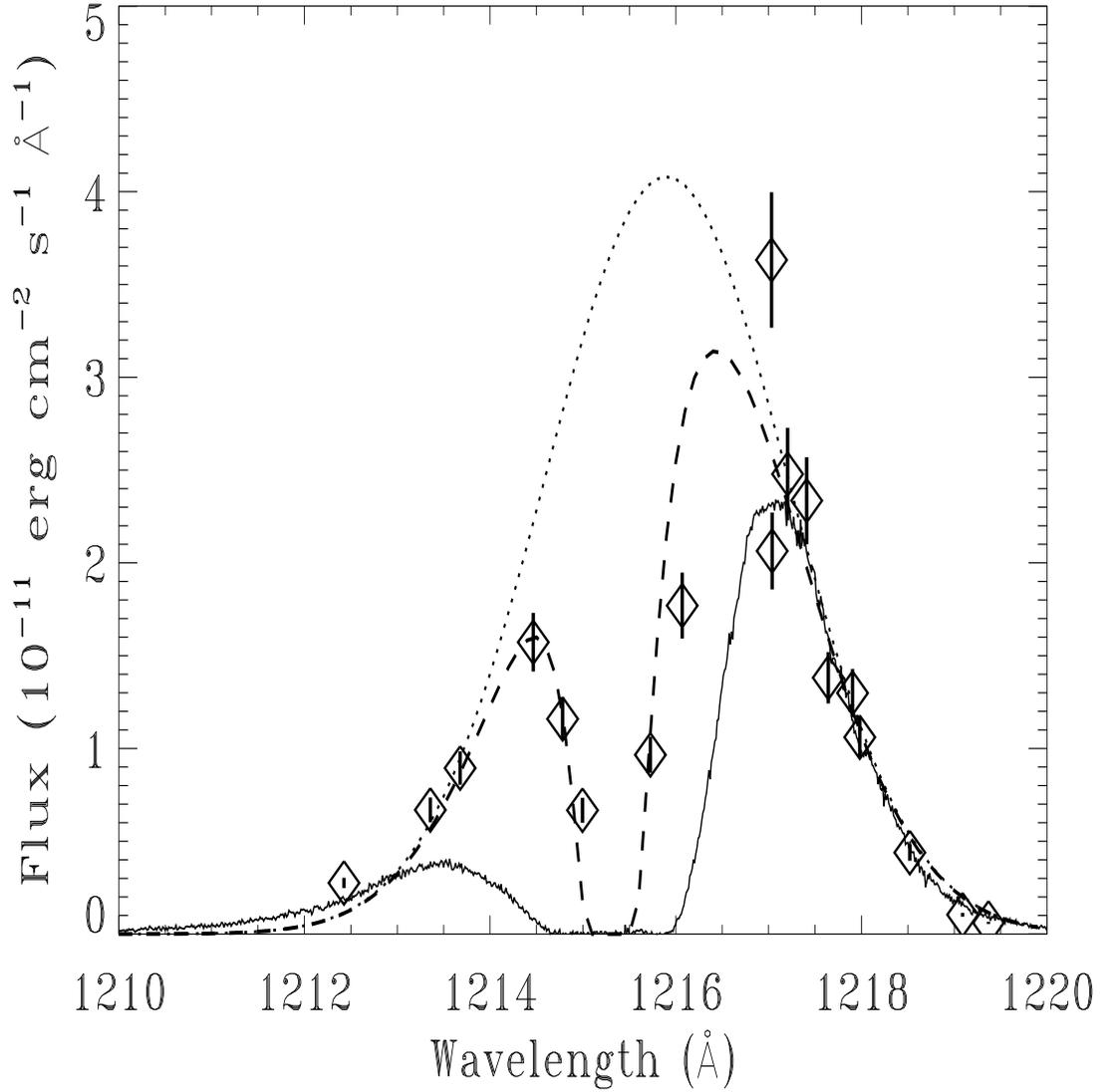}}
\label{fig:lya_best.eps}
\caption{The reconstructed \Lya\ profile at the wavelengths of the pumping
transitions (diamonds) for $T=2500$ K, $\log N($H$_2$)=18.5, and $\eta=0.25$ 
corresponds 
well to a plausible \Lya\ profile (dashed line), constructed using 
a single Gaussian profile (dotted line) and an absorption feature 
of $\log N$(\ion{H}{1})$=18.7$ blueshifted 90 \kms\ from line center.}
\end{figure}

\begin{figure}
\resizebox{5in}{8in}{\plotone{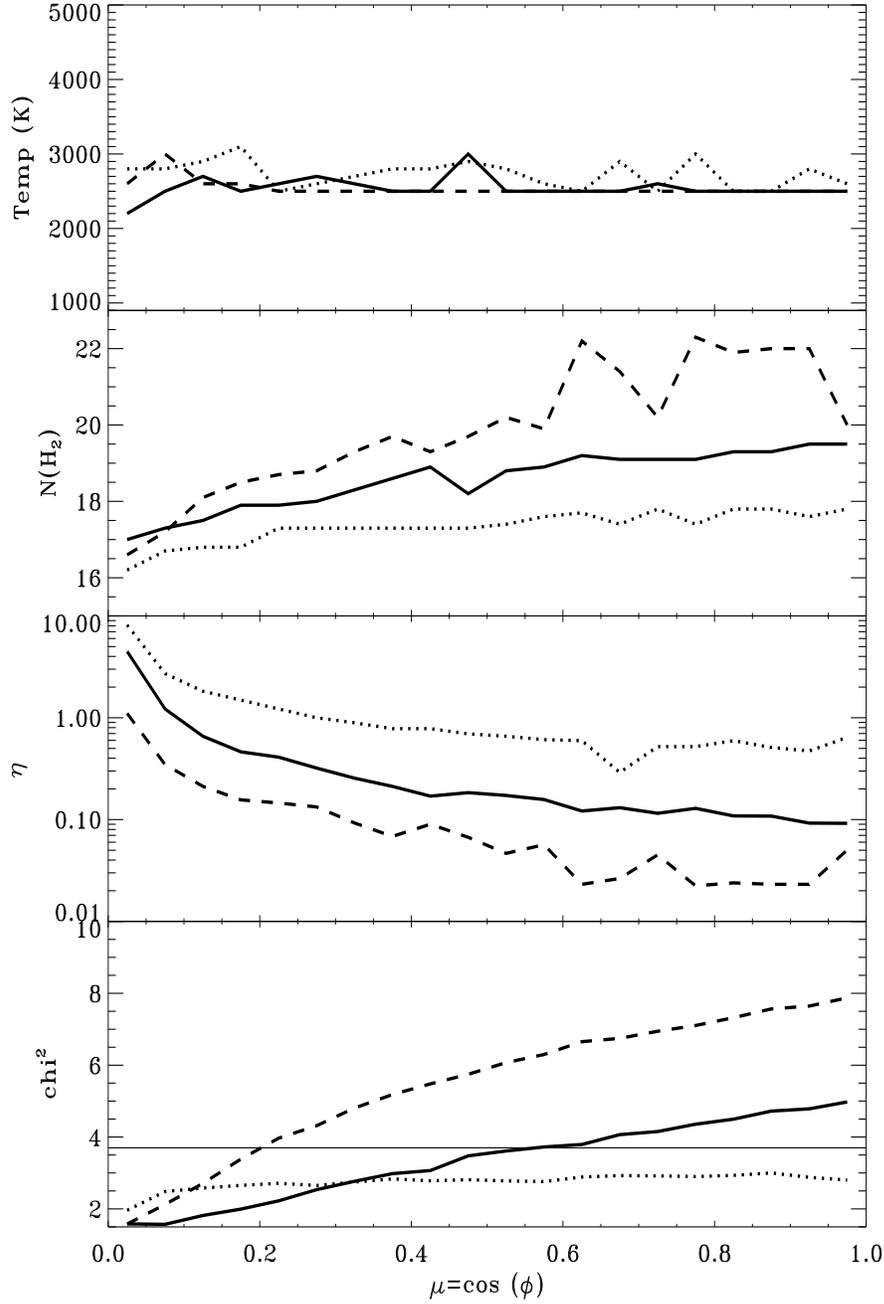}}
\label{fig:results_ang.eps}
\caption{Best-fit models to H$_2$ fluxes as a function of exit 
angle $\phi$ of H$_2$ photons, for the thick disk geometry (solid line), 
thin disk geometry (dashed line), and cloud geometry (dotted line).}
\end{figure}

\begin{figure}
\resizebox{5in}{8in}{\plotone{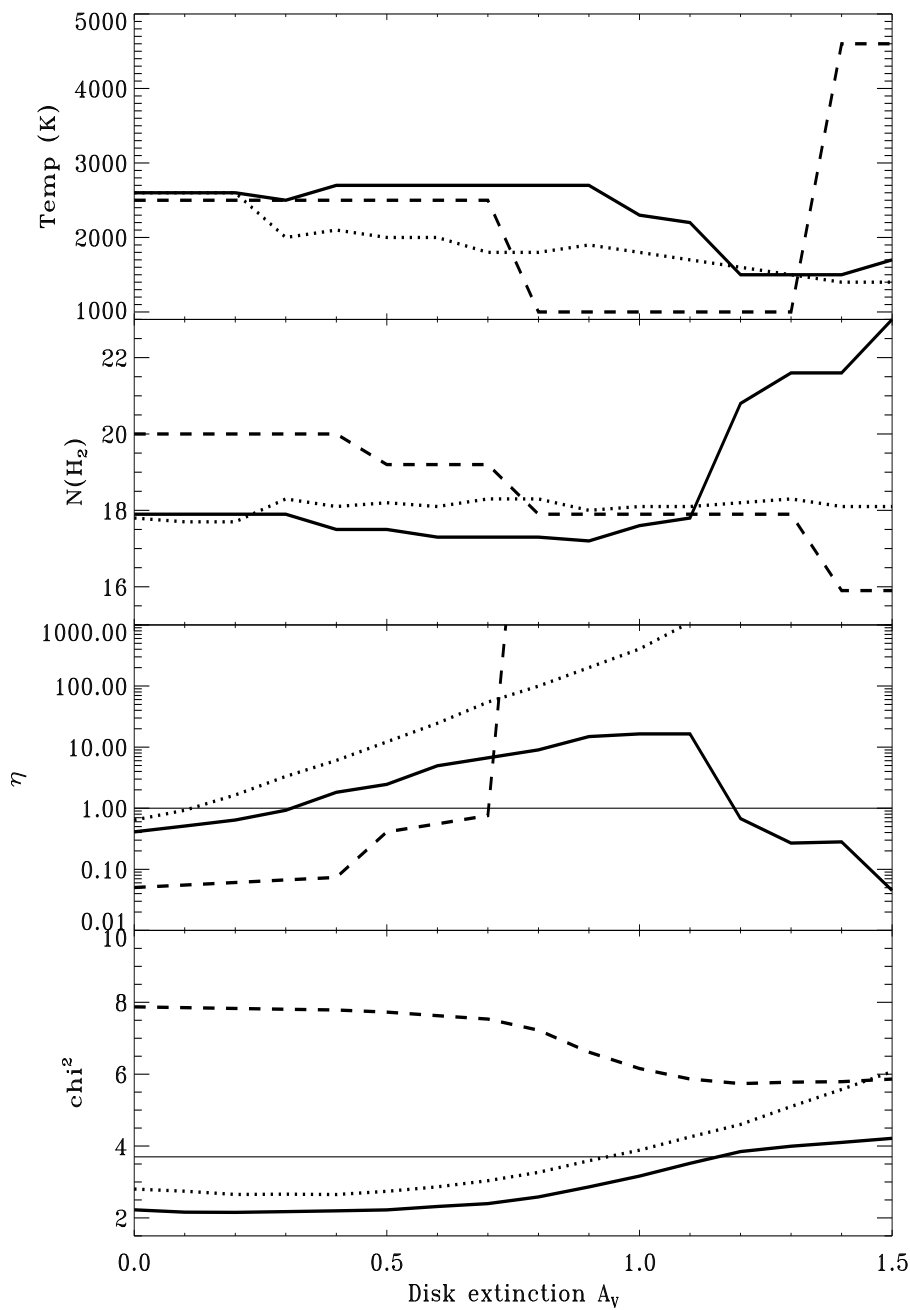}}
\label{fig:results_depth.eps}
\caption{Best-fit models to H$_2$ fluxes as a function of 
extinction within the H$_2$ slab, for the thick disk geometry (solid line), 
thin disk geometry (dashed line), and cloud geometry 
(dotted line).  The extinction is 
for a progression in which the average \Lya\ photon is absorbed 
at the midplane of the slab.}
\end{figure}

\pagebreak

\begin{figure}
\plotone{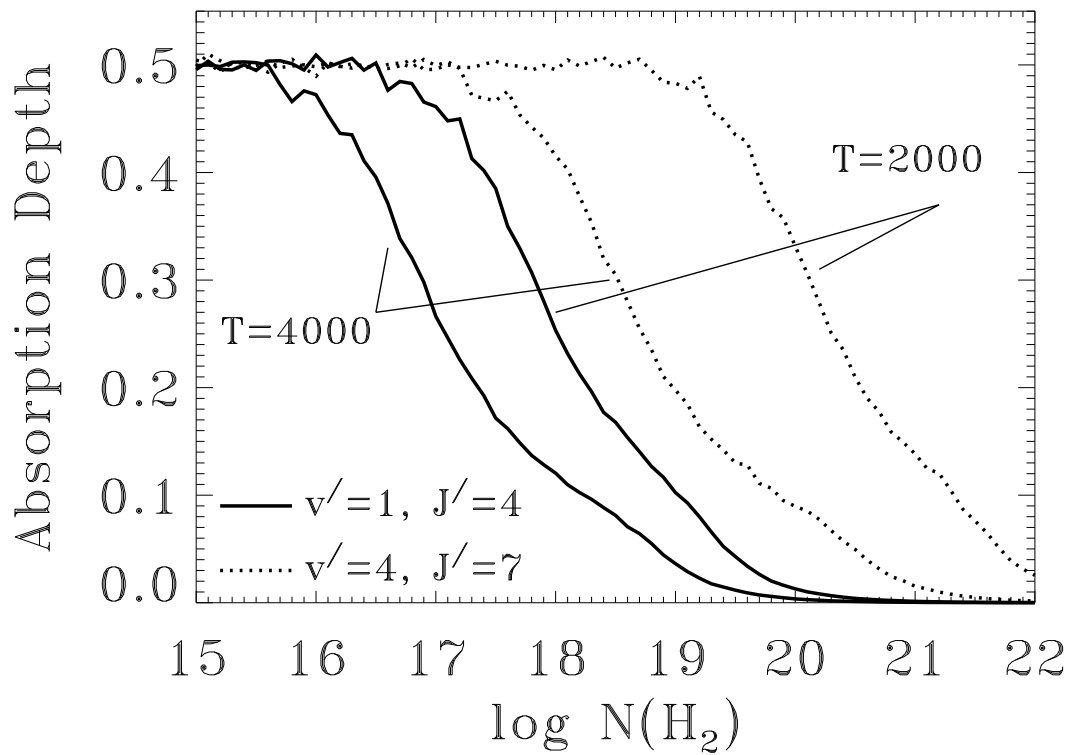}
\label{fig:depth.eps}
\caption{The percent of the slab depth that a \Lya\ photon excites H$_2$ to 
two different upper levels (absorption depth $d_{abs}$ in Table 2), at two
temperatures.  As the 
column density and temperature of the slab increase, the optical depth
 in the lower levels increase, and a typical \Lya\ photon is
 reprocessed into H$_2$ at a 
shallower depth.  This transition occurs at different column densities and 
temperatures for each progression, so that the reconstruction of 
the \Lya\ profile is sensitive to these parameters.}
\end{figure}

\clearpage
% Tables

\begin{table}
\label{tab:obs.tab}
\caption{LIST OF OBSERVATIONS}
\begin{tabular}{lccccccccccccc}
\hline
Star & Date & Instrument & Exposure Time & Grating & Aperture &
Wavelength Range \\
\hline
TW Hya & 2000 May 7 & \HST/\STIS &260  & G430L & $52\farcs \times0\farcs
0.2$ & 2900-5700\\
 & 2000 May 7 & \HST/\STIS & 1675 & E230M &  $0\farcs 2\times0\farcs 2$ & 2150--2950\\
 & 2000 May 7 & \HST/\STIS & 2300 & E140M & $0\farcs 5\times0\farcs 5$ & 1160--1700\\
 & 2000 Jun 22 & \FUSE & 2081 & $^1$ & $30\farcs \times 30 \farcs$ & 900-1185\\
V819 Tau & 2000 Aug 30 & \HST/\STIS & 360 &  G430L & $52\farcs \times0\farcs 1 $& 2900-5700\\
 & 2000 Aug 31 & \HST/\STIS & 2325 & E140M & $0\farcs 2\times0\farcs 2 $ & 2300-3100\\
 & 2000 Aug 30 & \HST/\STIS & 5700 & E140M & $0\farcs 2\times0\farcs 06 $ & 1160--1700\\
\hline
\multicolumn{7}{l}{$^1$The \FUSE\ instrument consists of several channels.}\\
\end{tabular}
\end{table}

\begin{table}
\label{tab:pumplist.tab}
\caption{H$_2$ PUMPING TRANSITIONS}
\rotatebox{90}{\begin{tabular}{lccccccccccccc}
\hline
Pump & $\lambda_{\rm calc}$ 
& \Elo$^1$  &  $A_{ul}^2$ & $d_{abs}^3$ &  EQW$_{slab}^4$ & $R_{unseen}^5$ & $P_{Dis}^6$ & $P_{Dis, mod}^7$ & $E_{dis}^8$ & $M_{dis}^9$ &
F$_{mod}^{10}$ \\
 & \AA & eV & $10^7$ s$^{-1}$ & \% & \AA & & & &eV \\
\hline
\hline
%3-2 P(4)$^6$&1174.923& 1.14  & - & - & -\\
1-1 P(11) & 1212.425 & 1.36  & 6.6  & 0.23  &  0.067  & 0.22  & 1.4(-5) & 0.0    & 0.07 & 0.0 & 4.5\\  
3-1 P(14) & 1213.356 & 1.79  & 10.1 & 0.43  &  0.028  & 0.38  & 0.015   & 0.018  & 0.10 & 2.5  & 5.9\\
4-2 R(12) & 1213.677 & 1.93  &  3.9 & 0.50  &  0.0092 & 0.30  & 0.050   & 0.067  & 0.15 & 4.4 & 2.7\\
3-1 R(15) & 1214.465 & 1.95  & 10.0 & 0.37  &  0.044  & 0.16  & 0.031   & 0.039  & 0.02 & 15.6  & 17\\
4-3 P(5) &  1214.781 & 1.65  & 5.5  & 0.43  &  0.030  & 0.25  & 1.9(-3) & 2.0(-3)& 0.11 & 0.0 & 11\\ 
4-3 R(6) & 1214.995  & 1.72  & 3.0  & 0.50  &  0.0092 & 0.68  & 9.2(-3)   & 0.011  & 0.07 & 0.5 & 2.4\\
1-2 R(6) & 1215.726  & 1.28  & 13.6 & 0.26  &  0.062  & 0.06  & 2.4(-6) & 0.0    & 0.08 & 0.0 & 16\\
1-2 P(5) & 1216.070  & 1.20  & 15.9 & 0.17  &  0.075  & 0.07  & 4.8(-7) & 0.0    & 0.06 & 0.0 & 33\\
3-3 R(2) & 1217.031  & 1.50  & 0.40 & 0.50  &  0.0015 & 0.45  & 2.8(-4) & 0.0    & 0.05 & 0.0 & 2.4\\
3-3 P(1) & 1217.038  & 1.48  & 1.7  & 0.50  &  0.0030 & 0.02  & 1.3(-4) & 0.0    & 0.06 & 0.0 & 2.9\\
0-2 R(0) & 1217.205  & 1.01  & 6.6  & 0.39  &  0.0413 & 0.09  & 4.3(-9) & 0.0    & 0.07 & 0.0 & 38\\
4-0 P(19) & 1217.410 & 2.21  & 4.4  & 0.47  &  0.014  & 0.47  & 0.42    & 0.51   & 0.18 & 64.0& 5.1\\
0-2 R(1) & 1217.643  & 1.02  & 7.8  & 0.21  &  0.0685 & 0.23  & 5.2(-9) & 0.0    & 0.07 & 0.0 & 32\\
2-1 P(13) & 1217.904 & 1.64  &  9.3 & 0.28  &  0.059  & 0.09  & 1.7(-3) & 2.9(-3)& 0.02 & 1.3 & 18\\
3-0 P(18) & 1217.982 & 2.03  & 3.2  & 0.50  &  0.0068 & 0.57  & 0.19    & 0.21   & 0.11 & 9.5 & 1.9\\
2-1 R(14) & 1218.521 & 1.79  & 7.6  & 0.43  &  0.028  & 0.42  & 5.9(-3) & 6.5(-3)& 0.09 & 0.6 & 3.6\\
0-2 R(2) & 1219.089  & 1.05  & 8.2  & 0.29  &  0.0576 & 0.47  & 6.6(-9) & 0.0    & 0.06 & 0.0 & 2.1\\ 
0-2 P(1) & 1219.368  & 1.02  & 20.1 & 0.28  &  0.0599 & 0.30  & 3.9(-9) & 0.0    & 0.06 & 0.0 & 1.2\\
0-5 P(18) & 1548.146 & 3.79  & 19.1 & 0.50  &  0.0    & 0.36  & 1.1(-4) & 0.0    & 0.11 & 0.0 & 3.0\\
\hline
\multicolumn{9}{l}{$^1$Lower energy level}\\
\multicolumn{9}{l}{$^2$Radiative decay coefficient \citep{Abg93}}\\
\multicolumn{9}{l}{$^3$Absorption depth of \Lya\ photon }\\
\multicolumn{9}{l}{$^4$Equivalent width of absorption profile within slab}\\
\multicolumn{9}{l}{$^5$Correction for unseen lines from upper level}\\
\multicolumn{9}{l}{$^6$Dissociation probability from upper state per
electronic excitation, $a(b)=a\times10^{b}$.}\\
\multicolumn{9}{l}{$^7$Dissociation probability from best-fit model.}\\
\multicolumn{9}{l}{$^8$Average kinetic energy released due to H$_2$ dissociation.}\\
\multicolumn{9}{l}{$^9$H$_2$ dissociation rate for best-fit model
($10^{-12} M_\odot$ yr$^{-1}$).}\\
\multicolumn{9}{l}{$^{10}$\Lya\ flux absorbed by lower level ($10^{-14}$ erg cm$^{-2}$ s$^{-1}$).}\\
%\multicolumn{9}{l}{$^11$Upper}\\
\end{tabular}}
\end{table}

\clearpage

\begin{table}
\label{tab:fuse.tab}
\caption{STRONGEST PREDICTED LYMAN H$_2$ LINES IN THE \FUSE\ BANDPASS}
\begin{tabular}{cccc}
\hline
ID & $\lambda_{calc}$ & $F_{mod}$ & $F_{obs}$\\
\hline
\hline
4-0 R(3) &   1053.973 &  1.1 & $<3$ \\
4-0 P(5) &   1065.593 &  1.6 & $<2$ \\
4-1 R(3) &   1101.889 &  2.4 & $<10$ \\
1-0 P(5) &   1109.311 &  1.0 & $<6$\\
3-1 P(1) &   1113.877 &  1.4 & $<6$\\
4-1 P(5) &   1113.949 &  1.4 & $<6$\\
1-1 R(3) &   1148.701 &  5.0 & $5.5(1.1)$\\
0-1 R(0) &   1161.693 &  1.9 & $13.1$(b)\\
1-1 P(5) &   1161.814 &  6.2 & $13.1$(b)\\
1-1 R(6) &   1161.949 &  1.8 & $13.1$(b)\\
0-1 R(1) &   1162.170 &  2.4 & $13.1$(b)\\
4-1 R(12) &   1164.596 &  1.1 & $<10$\\
2-0 P(13) &   1165.834 &  1.0 & $<6$\\
0-1 P(2) &   1166.255 &  2.6 & $<6$\\
0-1 P(3) &   1169.751 &  2.5 & $<10$\\
4-0 R(17) &   1176.325 &  1.0 & $<15$\\
3-1 R(12) &   1179.472 &  1.3 & $<10$\\
1-1 P(8) &   1183.309 &  2.7 & $<12$\\
2-1 R(11) &   1185.224 &  2.8 & $<12$\\
\hline
\multicolumn{4}{l}{~$F_{obs}$ refers to the observed fluxes and}\\
\multicolumn{4}{l}{~$F_{mod}$ to the model fluxes in units of}\\
\multicolumn{4}{l}{$10^{-15}$ erg cm$^{-2}$ s$^{-1}$,$1 \sigma$ error in ().}\\
\multicolumn{4}{l}{(b) indicates the line is blended with other H$_2$ lines.}\\
\end{tabular}
\end{table}

\begin{table}
\label{tab:undetect.tab}
\caption{UNDETECTED PROGRESSIONS}
\begin{tabular}{lccccccccccc}
\hline
Pump & $\lambda_{\rm calc}$ 
& $A_{ul}$ & \Elo\ & $P_{Dis}$ & $F_{\rm max}$
&$\lambda_{\rm max}$ &
$\lambda_{ref}$ & ID$_{ref}$\\
& \AA & $10^{-7}$ s$^{-1}$ & eV & & erg cm$^{-2}$ s$^{-1}$ & \AA & \AA & \\
\hline
\hline
1-1 R(12) & 1212.543 & 4.6 & 1.49 & 5(-5) & $3\times10^{-15}$ & 1578 & 1212.425 & 1-1 P(11)\\
7-4 P(5) & 1214.421 & 2.8 & 2.07 & 0.19 & $5\times10^{-15}$ & 1582 & 1214.465 & 3-1 R(15)\\
5-3 P(8) & 1218.575 & 6.6 & 1.89 & 0.04 &  $5\times10^{-15}$ & 1608 &
1218.521 & 2-1 R(14)\\ 
5-0 P(20) & 1217.716 & 5.4 & 2.39 & 0.48 &$2\times10^{-15}$ & 1223 &
1217.904 & 2-1 P(13)\\
2-2 P(8) & 1219.154 & 10.8 & 1.46 & 7(-5) &$5\times10^{-15}$ & 1579 &
1219.089 & 0-2 R(2)\\
2-2 R(9) & 1219.101 & 12.9 & 1.56 & 2(-4) &$5\times10^{-15}$ & 1572 & 1219.089 & 0-2 R(2)\\
5-3 R(9) & 1219.106 & 5.4 & 1.99 & 0.085 &$5\times10^{-15}$ & 1601 & 1219.089 & 0-2 R(2)\\
3-3 R(1) & 1215.541 & 0.46 & 1.48 & 2(-4) & $5\times10^{-15}$ & 1596 & 1215.726 & 1-2 R(6)\\
\hline
\end{tabular}
\end{table}

\begin{table}
\label{tab:results.tab}
\caption{BEST-FIT PARAMETERS FOR ALTERNATE GEOMETRIES}
\begin{tabular}{ccccccccccccccc}
\hline
\multicolumn{4}{c}{Geometry$^a$} && \multicolumn{4}{c}{Fit to H$_2$ line opacities} &&
\multicolumn{4}{c}{Fit to \Lya\ profile}\\
Number & Type & $\theta$ & $\phi$ & $A_V(D)$ & & $N($H$_2)$ & $T$ &
$\eta^b$ & $\chi^2_{\nu, H2}$ & & $N($H$_2)$ & $T$ & $\eta_{H2}$ & $\chi^2_{\nu,Ly\alpha}$\\
\hline
1 & Thick & 0 & 89 &  0  & & 17.0 & 2200 & 4.5 & 1.58      && 18.7 & 2400 & 0.29 & 0.9\\
2 & Thick &0 & 77 &  0   & & 17.9 & 2600 & 0.41& 2.22      && 18.5 & 2500 & 0.25 & 2.3\\
3 & Thick &0 & 77 &  0.5 & & 17.5 & 2700 & 2.5 & 2.22      && 18.8 & 2800 & 0.23 & 9.5\\
4 & Thick &0 & 77 &  1.0 & & 17.6 & 2300 & 16  & 3.16      && 18.7 & 5300 & 0.09 & 18\\
5 & Thick &0 & 77 &  1.5 & & 23.0 & 1700 & 0.05 & 4.21     && 19.0 & 5300 & 0.09 & 17\\
6 & Thick &0 & 43 &  0   & & 19.1 & 2600 & 0.12  & 4.15    && 18.2 & 2700 & 0.23 & 3.0\\
7 & Thick &0 & 0 &  0    & & 19.5 & 2500 & 0.09  & 4.98    && 18.4 & 2600 & 0.21 & 3.3\\
8 & Thin & 80 & 89  & 0  & & 16.6 & 2600 & 1.10  & 1.58    && 18.3 & 2200 & 0.31 & 1.1\\
9 & Thin & 80 & 77 &  0  & & 18.7 & 2500 & 0.15  & 3.97    && 17.9 & 2400 & 0.27 & 0.8\\
10 & Thin & 80 & 43  & 0  & & 20.2 & 2500 & 0.05  & 6.94    && 18.0 & 2400 & 0.23 & 0.8\\ 
11 &Thin & 80 & 0 &  0   & & 20.0 & 2500 & 0.05  & 7.87    && 18.0 & 2400 & 0.22 & 0.8\\
12 &Thin & 80 & 0 &  0.5 & & 19.2 & 2500 & 0.41   & 7.73   && 15.3 & 3300 & $10^5$ & 29\\
13 &Thin & 80 & 0 &  1.0 & & 17.9 & 1000 & $10^{13}$ & 6.16 && 19.3 & 4700 & 0.03 & 33\\
14 &Thin & 80 & 0 &  1.5 & & 15.9 & 4600 & $10^{12}$ & 5.86 && 19.4 & 4900 & 0.03 & 34\\
15 & Cloud & 80 & 89 &  0 & & 16.2 & 2800 & 8.2     & 1.96  && 18.5 & 2400 & 0.38 & 1.0\\
16 & Cloud & 80 & 77 &  0 & & 17.3 & 2500 & 1.2     & 2.71  && 18.7 & 2400 & 0.26 & 0.5\\
17 & Cloud & 80 & 43 &  0 & & 17.8 & 2500 & 0.52    & 2.92  && 18.7 & 2400 & 0.25 & 1.1\\
18 & Cloud & 80 & 0 &  0  & & 17.8 & 2600 & 0.64    & 2.80  && 18.8 & 2300 & 0.33 & 0.9\\
19 & Cloud & 80 & 0 &  0.5& & 18.2 & 2000 & 12       & 2.74 && 18.8 & 2400 & 3.32 & 1.4\\
20 & Cloud & 80 & 0 &  1.0& & 18.1 & 1800 & 400   & 3.88    && 18.8 & 2500 & 40.4 & 1.9\\
21 & Cloud& 80 & 0 &  1.5& & 18.1 & 1400 & 10$^4$ & 6.07   && 18.8 & 2700 & 400 & 2.8\\
\hline
\multicolumn{12}{l}{$^a$$\theta$ and $\phi$ are defined in \S3.2 and Fig. 8}\\
\multicolumn{12}{l}{$^b$$\eta_{H2}$ calculated from \Lya\ reconstruction
for parameters of best fit to H$_2$ fluxes.}\\
\end{tabular}
\end{table} 

\pagebreak

\begin{table}
\label{tab:fluxes.tab}
\caption{FLUXES OF STRONG LINES}
\begin{tabular}{ccc}
\hline
ID & $\lambda_{lab}$ & Flux$^1$\\
 & \AA\ & $10^{-12}$ erg cm$^{-2}$ s$^{-1}$\\ 
\hline
Total flux & 1170--1700 & $65$\\
\Lya$^2$ & 1215.67 & 44.7$^{4,5}$\\
\Lya$^3$ & 1215.67 & 80--160\\
Continuum & 1170--1700 & 6.0$^6$\\
H$_2$ (obs)& $<1700$ & 1.94$^{10}$\\
H$_2$ (mod)& $<1700$ & 2.20$^{11}$\\
C IV & 1548 & 1.86$^6$\\
He II & 1641& 1.35\\
C IV & 1551 & 0.98\\
C III & 977 & 0.44$^{4,8}$\\
C III & 1175 & 0.42\\
C III & 1175 & 0.35$^9$\\
O I & 1306 & 0.37 \\
O VI & 1032 & 0.31$^7$\\
N V & 1238 & 0.30 $^6$\\
O I & 1305 & 0.30$^{4}$\\
O I & 1302 & 0.28$^{4,5}$\\
O VI & 1038 & 0.15$^7$\\
C II & 1336 & 0.23$^{4}$\\
N V & 1243 & 0.12 \\
C II & 1335 & 0.11$^{4,5}$\\
Si IV & 1393&    0.076$^6$\\
Si IV & 1402&    0.034$^6$\\
O III] & 1666 & 0.023\\
\hline
\multicolumn{3}{l}{$^1$These fluxes are more accurate than those in Table 3.}\\
\multicolumn{3}{l}{$^2$Observed \Lya\ flux.}\\
\multicolumn{3}{l}{$^3$Flux estimated from reconstructed \Lya\ profile.}\\
\multicolumn{3}{l}{$^4$Wind absorption attenuates some blue emission.}\\
\multicolumn{3}{l}{$^5$IS/CS absorption absorbs some emission.}\\
\multicolumn{3}{l}{$^6$Flux from blended H$_2$ lines subtracted.}\\
\multicolumn{3}{l}{$^7$Average flux from \FUSE\ LiF1A, LiF2B channels.}\\
\multicolumn{3}{l}{$^8$Average flux from \FUSE\ SiC2A, SiC1B channels.}\\
\multicolumn{3}{l}{$^9$Average flux from \FUSE\ LiF2A, LiF1B channels.}\\
\multicolumn{3}{l}{$^{10}$Observed H$_2$ flux.}\\
\multicolumn{3}{l}{$^{11}$Total H$_2$ flux, including weak fluxes from
model.}\\
\end{tabular}
\end{table}

\end{document}